\documentclass{easychair}

\usepackage[final]{commenting}
\declareauthor{ri}{Radu}{blue}
\declareauthor{np}{Nicolas}{red}
\declareauthor{me}{Mnacho}{brown!70!black}

\usepackage{doc}

\usepackage{wrapfig}
\usepackage{pslatex}
\usepackage{amsfonts}
\usepackage{amsmath}
\usepackage{amssymb}
\usepackage{paralist}
\usepackage{hyperref}
\usepackage{qtree}
    \usepackage{caption}
    \DeclareCaptionLabelSeparator{none}{ }
\newcommand{\twoexptime}{\textsf{2-EXPTIME}\xspace}
\newcommand{\elementary}{\textsf{ELEMENTARY}\xspace}

\newcommand{\aexpspace}{\textsf{AEXPSPACE}\xspace}
\newcommand{\coaexpspace}{\textsf{co-AEXPSPACE}\xspace}

\newif\ifLongVersion\LongVersiontrue

 \newtheorem{theorem}{Theorem}
\newtheorem{definition}[theorem]{Definition}
 \newtheorem{lemma}[theorem]{Lemma}
 \newtheorem{proposition}[theorem]{Proposition}

\newtheorem{example}[theorem]{Example}

\newcommand{\bigO}{\mathcal{O}} 
\newcommand{\np}{$\mathsf{NP}$}
\newcommand{\exptime}{$\mathsf{EXPTIME}$}

\newcommand{\set}[1]{\left\{ #1 \right\}}

\renewcommand{\vec}[1]{\mathbf #1}

\newcommand{\len}[1]{{|{#1}|}}
\newcommand{\card}[1]{{||{#1}||}}

\newcommand{\arrow}[2]{\xrightarrow{{\scriptscriptstyle #1}}_{{\scriptscriptstyle #2}}}

\newcommand{\nat}{{\bf \mathbb{N}}}

\newcommand{\proj}[2]{{#1}\!\!\downarrow_{{#2}}}

\renewcommand{\paragraph}[1]{\noindent{\bf #1}}

\let\Asterisk\undefined
\newcommand{\Asterisk}{\mathop{\scalebox{1.7}{\raisebox{-0.2ex}{$\ast$}}}}%

\newif\ifLongVersion\LongVersiontrue
\renewcommand{\proof}[1]{\ifLongVersion \noindent\emph{Proof}: {#1} \else\fi}

\newcommand{\locs}{\mathsf{Loc}}

\newcommand{\nil}{\mathsf{nil}}
\newcommand{\seplog}{\mathsf{SL}}

\newcommand{\seplogk}[1]{\seplog^{\!\scriptstyle{#1}}}

\newcommand{\fv}[1]{\mathsf{fv}({#1})}

\newcommand{\dom}{\mathrm{dom}}

\newcommand{\interv}[2]{\llbracket#1\mathrel{{.}\,{.}}\nobreak#2\rrbracket}

\newcommand{\img}{\mathrm{rng}}

\newcommand{\vars}{\mathsf{Var}}
\newcommand{\preds}{\mathsf{Pred}}

\renewcommand{\int}{\mathsf{\scriptscriptstyle{i}}}

\newcommand{\astore}{\mathfrak{s}}
\newcommand{\aheap}{\mathfrak{h}}

\renewcommand{\iff}{\Leftrightarrow}

\newcommand{\isdef}{\stackrel{\scalebox{.9}{$\scriptscriptstyle{\mathsf{def}}$}}{=}}

\newcommand{\finmap}{\rightharpoonup_{\mathit{fin}}}
\newcommand{\terms}{\mathsf{Term}}
\newcommand{\rank}{\kappa}
\newcommand{\seq}{\approx}
\newcommand{\sneq}{\not \seq}
\newcommand{\nloc}{\mathit{nil}}

\newcommand{\asys}{\mathcal{S}}
\newcommand{\arule}{\rho}

\newcommand{\nodes}{\texttt{nodes}}

\newcommand{\utrees}[2]{\mathcal{T}_{#2}({#1})}
\newcommand{\charform}[1]{\Upsilon({#1})}

\newcommand{\plab}{\Lambda}

 \usepackage{stmaryrd}
 \usepackage{xspace}
 \usepackage{amsmath}
 \usepackage{txfonts}
 \usepackage{mfirstuc}
 \usepackage{graphicx}
 \usepackage{paralist}
 \usepackage{ifthen}
 \usepackage{scalerel}

\newcommand{\rep}[3]{#1[#2 \leftarrow #3]}  

\renewcommand{\vec}[1]{\boldsymbol{#1}}   

\newcommand{\leftmv}{\leftarrow}
\newcommand{\rightmv}{\rightarrow}

\newcommand{\blank}{\textsc{b}}
\newcommand{\amove}{\mu}
\newcommand{\atm}{ATM\xspace}

\newcommand{\blind}{\bullet}
\newcommand{\nn}{\mathfrak{N}}

\newcommand{\allcst}{\mathsf{Const}}
\newcommand{\act}[1]{\overline{#1}}
\newcommand{\movenc}[1]{\widetilde{#1}}
\newcommand{\psymfunc}{\Delta}
\newcommand{\psym}[2]{\psymfunc_{#1}({#2})}
\newcommand{\encodes}[3]{{#1} \rhd_{#2} {#3}}
\newcommand{\actencodes}[3]{{#1} \rhd_{#2} {#3}}
\newcommand{\bin}[1]{\mathrm{bin}({#1})}
\newcommand{\transet}[1]{\tau({#1})}

\newcommand{\mv}[2]{#1^{#2}}				
\newcommand{\step}[1]{\arrow{#1}{}}

\newcommand{\nt}{\mathfrak{B}}

\newcommand{\ground}{predicate-free\xspace}

\newcommand{\goal}{head\xspace}

\newcommand{\labeling}{predicate decoration\xspace}

\newcommand{\shortversion}[2]{#1}
\renewcommand{\shortversion}[2]{#2} 

\begin{document}

\title{Entailment Checking in Separation Logic with Inductive
  Definitions is \twoexptime-hard}

\titlerunning{Entailment Checking in Separation Logic with Inductive
  Definitions is \twoexptime-hard}

\author{Mnacho Echenim\inst{1}, Radu Iosif\inst{2} and Nicolas Peltier\inst{1}}

\authorrunning{Echenim, Iosif and Peltier}


\institute{Univ. Grenoble Alpes, CNRS, LIG, F-38000 Grenoble France
\and
Univ. Grenoble Alpes, CNRS, VERIMAG, F-38000 Grenoble France}

\maketitle

\begin{abstract}
The entailment between separation logic formul{\ae} with inductive
predicates, also known as symbolic heaps, has been shown to be
decidable for a large class of inductive definitions
\cite{IosifRogalewiczSimacek13}. Recently, a \twoexptime\ algorithm
was proposed \cite{KatelaanMathejaZuleger19,ZK20} and an \exptime-hard
bound was established in \cite{DBLP:conf/atva/IosifRV14}; however no
precise lower bound is known. In this paper, we show that deciding
entailment between predicate atoms is \twoexptime-hard. The proof is
based on a reduction from the membership problem for exponential-space
bounded alternating Turing machines \cite{ChandraKozenStockmeyer81}.
\end{abstract}

\section{Introduction}

Separation logic is a particular case of the logic of bunched
implications \cite{DBLP:journals/bsl/OHearnP99}. It was introduced in
\cite{Reynolds02} as an extension of Hoare logic intended to
facilitate reasoning on mutable data-structures, and it now forms the
basis of highly successful static analyzers such as, e.g., Infer
\cite{DBLP:conf/nfm/CalcagnoDDGHLOP15}, SLAyer
\cite{DBLP:conf/cav/BerdineCI11} or Predator
\cite{DBLP:conf/cav/DudkaPV11}.
The assertions in this logic describe {\em heaps}, that are finite
partial functions mapping locations to tuples of locations (records),
intended to model dynamically allocated objects.  The usual
connectives of propositional logic are enriched with a special
connective, called the {\em separating conjunction}, that permits to
assert that two formul{\ae} hold on disjoint parts of the heap,
allowing for more concise and more natural specifications.  In this
paper, we consider the fragment of separation logic formul{\ae} known
as {\em symbolic heaps}, consisting of separated conjunctions of
atoms. Such atoms may be equational atoms, asserting equalities or
disequalities between memory locations; points-to atoms asserting that
some location refers to a given record; or may be built on additional
predicates that assert that a part of the memory has some specific
shape (such as a tree).  For genericity, such predicates are
associated with user-provided inductive definitions that allow one to
describe custom data-structures.  For example, the formula $x \mapsto
(y,z) * p(y)$ states that the heap is composed of two disjoint parts:
a first location $x$ pointing to a tuple of locations $(y,z)$ and a
second part described by $p(y)$.  Given the inductive definition:
\[p(x) \Leftarrow x \mapsto (\nil,\nil) \qquad p(x) \Leftarrow \exists y_1,y_2~.~ x \mapsto (y_1,y_2) * p(y_1) * p(y_2)\]
$p(y)$ states that the considered part of the heap is a \comment[np]{footnote added} binary tree\footnote{For conciseness we omit the rules for the two cases where one of the children is $\nil$ but the other one is not.} the  rooted at  $y$.

This logic provides a very convenient way to describe graph-like
data-structures.  Satisfiability is \exptime-complete for such
formul{\ae} \cite{DBLP:conf/csl/BrotherstonFPG14}, but entailment is
not decidable in general\footnote{Entailment does not reduce to
  satisfiability since the considered logic has no negation.}
\cite{DBLP:conf/atva/IosifRV14,AntonopoulosGorogiannisHaaseKanovichOuaknine14}.
However, the entailment problem was proven to be decidable for a large
class of inductive definitions, with syntactical restrictions that
ensure the generated heap structures have a bounded-tree width
\cite{IosifRogalewiczSimacek13}, using a reduction to monadic second-order logic interpreted over graphs. An \exptime-hard bound was
established in \cite{DBLP:conf/atva/IosifRV14}, and very recently, a
\twoexptime\ algorithm has been proposed. \comment[np]{modifs} Although the algorithm in
\cite{KatelaanMathejaZuleger19} (implemented in the system {\sc Harrsh}) 
is practically successful (as evidenced by the 
experimental results reported in \cite{KatelaanMathejaZuleger19} and at \url{https://github.com/katelaan/harrsh}), it was
discovered in \cite{ZK20} that it was incomplete, and some techniques
are proposed to fix this issue \comment[np]{to Radu: should we cite here their LPAR 2020 paper?} (a complete description of the new
algorithm is available in the technical report \cite{PMZ20}).  In this
paper, we show that the problem is \twoexptime-hard, even if only
entailment between predicate atoms is considered. The proof relies on
a reduction from the membership problem for alternating Turing
machines \cite{ChandraKozenStockmeyer81} whose working tape is
exponentially bounded in the size of the input. This result gives the
tight complexity for the problem, whose upper bound is
\twoexptime\ \cite{KatelaanMathejaZuleger19,PMZ20}. 

This paper is a thoroughly revised version of a paper that was
presented at the workshop ADSL 2020 (with no formal proceedings).
\shortversion{Due to space restrictions, the proofs of several technical lemmas are \comment[np]{modifs} omitted.
They can be found in \cite{longversion}.}{}


\section{Separation Logic with Inductive Definitions}


For any set $S$, we denote by $\card{S} \in \nat \cup \{ \infty \}$ its
cardinality.  For a partial mapping $f : A \rightharpoonup B$, let
$\dom(f) \isdef \set{x \in A \mid f(x) \text{\ is defined}}$ and $\img(f) \isdef
\set{f(x) \mid x \in \dom(f)}$ be its domain and range, respectively, and we
write $f : A \finmap B$ if $\card{\dom(f)} < \infty$. Given integers
$n,m$, we denote by $\interv{n}{m}$ the set $\set{n,n+1, \ldots,
  m}$ (with $\interv{n}{m} = \emptyset$ if $n > m$). 
By a slight abuse of notation, we write $t \in
\vec{t}$ if $\vec{t} = (t_1, \ldots, t_n)$ and $t=t_i$, for some $i \in \interv{1}{n}$.

Let $\vars = \set{x,y,\ldots}$ be an infinite countable set of {\em
  variables} and $\preds = \set{p,q,\ldots}$ be an infinite countable
set of uninterpreted relation symbols, called \emph{predicates}. Each
predicate $p$ has an arity $\#p \geq 1$, denoting the number of its
arguments. In addition, we consider a special function symbol $\nil$,
of arity zero. A \emph{term} is an element of the set $\terms \isdef
\vars \cup \set{\nil}$. Let $\rank\geq1$ be an integer constant fixed
throughout this paper, intended to denote the number of record fields.
The logic $\seplogk{\rank}$ is the set of formul{\ae} generated
inductively as follows: \comment[np]{removed emp which is useless}
\[\begin{array}{rcl}
\phi & := & t_0 \mapsto (t_1, \ldots, t_\rank) \mid p(t_1,
\ldots, t_{\#p}) \mid t_1 \seq t_2 \mid t_1 \sneq t_2 \mid \phi_1 * \phi_2
\mid \exists x ~.~ \phi_1
\end{array}\]
where $p \in \preds$, $t_i \in \terms$, for all $i \in
\interv{0}{\max(\rank,\#p)}$ and $x \in \vars$. A \emph{\ground
  formula} is a formula of $\seplogk{\rank}$ in which no predicates
occur. A formula of the form $t_0 \mapsto (t_1, \ldots, t_\rank)$
[resp.\ $p(t_1, \ldots, t_{\#p})$] is called a \emph{points-to atom}
[resp.\ \emph{predicate atom}]. We write $\fv{\phi}$ for the set of
\emph{free} variables in $\phi$, i.e., the variables $x$ that occur in
$\phi$ outside of the scope of any existential quantifier $\exists
x$. 
If $\fv{\phi} = \{ x_1,\dots,x_n \}$ then $\phi[y_1/x_1, \ldots,
  y_n/x_n]$ denotes the formula obtained from $\phi$ by simultaneously
substituting each $x_i$ with $y_i$, for $i \in \interv{1}{n}$. \comment[np]{removed: A
\emph{substitution} is a mapping $\sigma : \vars \rightarrow \terms$
and we denote by $\phi\sigma$ the formula $\phi[\sigma(x_1)/x_1,
  \ldots,\sigma(x_n)/x_n]$, where $\fv{\phi}=\set{x_1,\ldots,x_n}$.}

To interpret $\seplogk{\rank}$ formul{\ae}, we consider a fixed,
countably infinite set $\locs$ of \emph{locations} and a designated
location $\nloc \in \locs$. The semantics of $\seplogk{\rank}$
formul{\ae} is defined in terms of \emph{structures} $(\astore,
\aheap)$, where: \begin{compactitem}
  \item $\astore : \terms \rightarrow \locs$ is a total mapping of
    terms into locations, called \emph{store}, such that
    $\astore(\nil) = \nloc$,
  \item $\aheap : \locs \finmap \locs^\rank$ is a finite partial
    mapping of locations into $\rank$-tuples of locations, called
    \emph{heap}, such that $\nloc \not\in \dom(\aheap)$. 
\end{compactitem}
A location is {\em allocated} in a heap $\aheap$ if it occurs in
$\dom(\aheap)$.  Two heaps $\aheap_1$ and $\aheap_2$ are
\emph{disjoint} iff $\dom(\aheap_1) \cap \dom(\aheap_2) = \emptyset$,
in which case their \emph{disjoint union} is denoted by $\aheap_1
\uplus \aheap_2$, undefined if $\dom(\aheap_1) \cap \dom(\aheap_2)
\neq \emptyset$.

The satisfaction relation $\models$ between structures and
\ground $\seplogk{\rank}$ formul{\ae}  is defined, as usual, recursively on the
syntax of formul{\ae}:
\[\begin{array}{rclcl}
(\astore, \aheap) & \models & t_1 \seq t_2 & \iff &
\aheap = \emptyset \text{ and } \astore(t_1) = \astore(t_2) \\
(\astore, \aheap) & \models & t_1 \sneq t_2 & \iff &
\aheap = \emptyset \text{ and } \astore(t_1) \neq \astore(t_2) \\
(\astore, \aheap) & \models & t_0 \mapsto (t_1, \ldots, t_\rank) & \iff &
\dom(\aheap) = \set{\astore(t_0)} \text{ and } \aheap(\astore(t_0)) = (\astore(t_1), \ldots, \astore(t_\rank)) \\

(\astore, \aheap) & \models & \phi_1 * \phi_2 & \iff &
\text{there are disjoint heaps $\aheap_1$ and $\aheap_2$, such that $\aheap = \aheap_1 \uplus \aheap_2$} \\
&&&& \text{and $(\astore, \aheap_i) \models \phi_i$, for each $i=1,2$} \\
(\astore, \aheap) & \models & \exists x ~.~ \phi & \iff & (\astore[x \leftarrow \ell], \aheap) \models \phi,
\text{ for some $\ell \in \locs$,}
\end{array}\]
where $\astore[x \leftarrow \ell]$ is the store mapping $x$ into
$\ell$ and behaving like $\astore$ for all $t \in
\terms\setminus\set{x}$.  Note that the semantics of $t_1\seq t_2$ and
$t_1\sneq t_2$ is \emph{strict}, meaning that these atoms are
satisfied only if the heap is empty\footnote{This semantics avoids
  using boolean conjunction: $\phi \wedge x = y \iff \phi * x \seq y$,
  where $x=y$ iff $x$ and $y$ are assigned the same location.}.

\subsection{Unfolding Trees}

We now extend the previous semantics to handle formul{\ae} containing
predicate atoms.  We assume that such predicates are associated with a
set $\asys$ of \emph{rules} of the form $p(x_1, \ldots, x_{\#p})
\Leftarrow \arule$, where $\arule$ is an $\seplogk{\rank}$ formula
such that $\fv{\arule} \subseteq \{x_1, \ldots, x_{\#_p}\}$. We refer
to $p(x_1, \ldots, x_{\#p})$ as the \emph{\goal}, and to $\arule$ as
the \emph{body} of the rule. A rule is a \emph{base rule} if its body
is a \ground formula. We write $p(x_1,\ldots,x_{\#p}) \Leftarrow_\asys
\arule$ if the rule $p(x_1,\ldots,x_{\#p}) \Leftarrow \arule$ belongs
to $\asys$. In this section, we consider a given set of rules $\asys$.

The above semantics is extended to formul{\ae} that are not \ground,
by recursively replacing predicate symbols by the body of a defining
rule until a simple formula is obtained, in a finite number of
steps. For technical convenience, we place the steps of an unfolding
sequence in a tree, such that the descendants of a node represent the
unfoldings of predicate atoms produced by the unfolding of that
particular node. Formally, a {\em tree} $t$ is defined by a set of
nodes $\nodes(t)$ and a function mapping each node $w\in \nodes(t)$ to
its {\em label}, denoted by $t(w)$.  The set $\nodes(t)$ is a finite
prefix-closed subset of $\nat^*$, where $\nat^*$ is the set of finite
sequences of non-negative  integers, meaning that if $w$ and $wi$ are
elements of $\nodes(t)$ for some $i \in \nat \setminus \set{0}$, then
so is $wj$ for all $j\in\interv{0}{i-1}$.  We write $\len{w}$ for the
length of the sequence $w$ and $\lambda$ for the empty sequence (so
that $\len{\lambda}=0$). The \emph{root} of $t$ is $\lambda$, the {\em
  children} of a node $w\in \nodes(t)$ are the nodes $wi \in
\nodes(t)$, where $i \in \nat$, and the {\em parent} of a node $wi$
with $i \in \nat$ is $w$ (hence, $\lambda$ has no parent).  The
subtree of $t$ rooted at $w$ is denoted by $\proj{t}{w}$; it is
formally defined by $\nodes(\proj{t}{w}) \isdef \{ w' \mid ww' \in
\nodes(t) \}$ and $\proj{t}{w}(w') \isdef t(ww')$, for all $w' \in
\nodes(\proj{t}{w})$. For simplicity, we define unfolding trees below
only for predicate atoms\footnote{An unfolding tree for a generic
  $\seplogk{\rank}$ formula can be obtained by joining the unfolding
  trees of its predicate atoms under a common root.}:

\begin{definition}\label{def:unfolding-tree}
  An \emph{unfolding tree} of a predicate atom $p(t_1,\ldots,t_{\#p})$
  is a tree $u$, such that, for all $w \in \nodes(u)$, we have $u(w) =
  (q(s_1,\ldots,s_{\#q}),\psi)$, for a predicate atom
  $q(s_1,\ldots,s_{\#q})$ and a formula $\psi$,
  where: \begin{compactenum}
  \item\label{it1:unfolding-tree} if $w = \lambda$ then $q(s_1,\ldots,s_{\#q}) = p(t_1, \ldots,
    t_{\#p})$,
  \item\label{it2:unfolding-tree} $\psi = \arule[s_1/x_1,\ldots,s_{\#q}/x_{\#q}]$, for a rule
    $q(x_1,\ldots,x_{\#q}) \Leftarrow_\asys \arule$, and
  \item\label{it3:unfolding-tree} there exists a bijective mapping
    from the set of occurrences of predicate atoms in $\psi$ and the
    children\footnote{In particular, $\psi$ is a \ground formula iff
      $w$ is a leaf.} of $w$, such that if an atom $r(v_1, \ldots,
    v_{\#r})$ is mapped to $wi$, for some $i \in \nat$, then $u(wi)$
    is of the form $(r(v_1, \ldots, v_{\#r}),\psi_i)$, for some
    formula $\psi_i$.
  \end{compactenum}
  We denote by $\utrees{p(t_1,\ldots,t_{\#p})}{\asys}$ the set of
  unfolding trees for $p(t_1,\ldots,t_{\#p})$.
\end{definition}
Given an unfolding tree $u \in \utrees{p(t_1,\ldots,t_{\#p})}{\asys}$,
such that $u(\lambda)=(p(t_1,\ldots,t_{\#p}),\psi)$, we define its
\emph{characteristic formula} inductively, as the \ground formula
$\charform{u}$ obtained from $\psi$ by replacing each occurrence of an
atom $q(s_1, \ldots, s_{\#q})$ by $\charform{\proj{u}{i}}$, where $i$
denotes the child of $w$ to which $q(s_1, \ldots, s_{\#q})$
 is
mapped\footnote{Note that the bijection between atoms and children is
  not necessarily unique. However, it is easy to check that all these
  mappings will eventually yield the same formula, up to a permutation
  of atoms.} by the bijection of point (\ref{it3:unfolding-tree}) in
Definition \ref{def:unfolding-tree}.  More
precisely, if $\psi = \exists y_1 \ldots \exists y_n ~.~ \varphi *
\Asterisk_{i=1}^m q_i(s^i_1, \ldots, s^i_{\#q_i})$, where $\varphi$ is
\ground, then $\charform{u} = \exists y_1 \ldots \exists y_n ~.~
\varphi * \Asterisk_{i=1}^m \charform{\proj{u}{i}}$.

Given an $\seplogk{\rank}$ formula $\phi$ and a structure
$(\astore,\aheap)$, we write $(\astore,\aheap) \models_\asys \phi$ if
and only if $(\astore,\aheap) \models \psi$ \comment[np]{modif} for some formula $\psi$ is obtained
from $\phi$ by syntactically replacing each 
occurrence of a predicate atom $p(t_1,\ldots,t_{\#p})$ in $\phi$ with
a formula $\charform{u}$, for some unfolding tree $u \in
\utrees{p(t_1,\ldots,t_{\#p})}{\asys}$. A structure $(\astore,\aheap)$
such that $(\astore,\aheap) \models_\asys \phi$ is called an
$\asys$-model of $\phi$, or simply a model of $\phi$, when $\asys$ is
clear from the context. 

We may now define the class of entailment
problems, which are the concern of this paper:

\begin{definition}\label{def:entailment}
  Given a set of rules $\asys$ and two $\seplogk{\rank}$ formul{\ae}
  $\phi$ and $\psi$, is it the case that every $\asys$-model of $\phi$
  is an $\asys$-model of $\psi$? Instances of the entailment problem
  are denoted  $\phi \models_\asys \psi$.
\end{definition}

\section{A Decidable Class of Entailments}

\label{sect:fragment}

In general, the entailment problem is undecidable
\cite{DBLP:conf/atva/IosifRV14,AntonopoulosGorogiannisHaaseKanovichOuaknine14}.
Thus we consider a subclass of entailments for which decidability
(with elementary recursive complexity) was proved in
\cite{IosifRogalewiczSimacek13} and provide a \twoexptime\ lower bound
for this problem.  The decidable class is defined by three
restrictions on the rules used for the interpretation of predicates,
namely \emph{progress}, \emph{connectivity} and \emph{establishment},
recalled next.

First, the \emph{progress} condition requires that each rule adds to
the heap exactly one location, namely the one associated with the first
parameter of the \goal. Second, the \emph{connectivity} condition
requires that all locations added during an unfolding of a predicate
atom $p(\vec{t})$ form a connected tree-like structure.

\begin{definition}\label{def:progress-connectivity}
  A set of rules $\asys$ is \emph{progressing} if and only if the body
  $\arule$ of each rule $p(x_1, \ldots, x_{\#p}) \Leftarrow_\asys
  \arule$ is of the form $\exists z_1 \ldots \exists z_m ~.~ x_1
  \mapsto (y_1, \ldots, y_\rank) * \psi$ and $\psi$ contains no
  occurrence of a points-to atom. If, moreover, each occurrence of a
  predicate atom in $\psi$ is of the form $q(y_i, u_1, \ldots,
  u_{\#q-1})$, for some $i \in \interv{1}{\rank}$, then $\asys$ is
  \emph{connected}.
\end{definition}

The progress and connectivity conditions induce a tight relationship between the
models of predicate atoms and their corresponding unfolding trees,
formalized below:

\begin{definition}\label{def:embedding}
  Given a heap $\aheap$ and a tree $t$, an \emph{embedding of $t$
  into $\aheap$} is a bijection $\plab : \nodes(t) \rightarrow
  \dom(\aheap)$ such that $\plab(wi) \in \aheap(\plab(w))$, for each
  node $wi \in \nodes(t)$, where $i \in \nat$.
\end{definition}

The following lemma states that every unfolding tree of a predicate
atom can be embedded into the heap of a model of its characteristic
formula.

\begin{lemma}\label{lemma:labeling}
  Let $\asys$ be a progressing and connected set of rules and
  $(\astore,\aheap)$ be a structure such that $(\astore,\aheap)
  \models_\asys p(t_1,\ldots,t_{\#p})$. Then there exists an unfolding
  tree $u \in \utrees{p(t_1,\ldots,t_{\#p})}{\asys}$ such that
  $(\astore,\aheap)\models\charform{u}$, and an embedding of $u$ into
  $\aheap$.
\end{lemma}
\shortversion{}{\proof{
If $(\astore,\aheap) \models_\asys p(t_1,\ldots,t_{\#p})$,
then there exists $u \in \utrees{p(t_1,\ldots,t_{\#p})}{\asys}$,
such that $(\astore,\aheap) \models \charform{u}$, by the definition
of $\models_\asys$. The embedding $\plab$ is built inductively on
the structure of $u$, as follows: \begin{compactitem}
\item If $\nodes(u) = \set{\lambda}$ (we have assumed that trees are
  nonempty) then, because $\asys$ is progressing, we must have
  \(u(\lambda) = (p(t_1,\ldots,t_{\#p}), \exists z_1 \ldots \exists
  z_n ~.~ t_1 \mapsto (t'_1,\ldots,t'_\rank) * \psi)\), where $\psi$
  is a separating conjunction of equational atoms. Then we obtain
  $\charform{u} = \exists z_1 \ldots \exists z_n ~.~ t_1 \mapsto
  (t'_1,\ldots,t'_\rank) * \psi$ and, because $(\astore,\aheap)
  \models \charform{u}$, we have $\dom(\aheap) =
  \set{\astore(t_1)}$. In this case, we define $\plab =
  \set{(\lambda,\astore(t_1))}$, which is a bijection between
  $\nodes(u)$ and $\dom(\aheap)$. Moreover, $\plab$ is an embedding,
  since $u$ contains only one node.
\item Otherwise, \(u(\lambda) = (p(t_1,\ldots,t_{\#p}), \exists y_1
  \ldots \exists y_n ~.~ t_1 \mapsto (t'_1,\ldots,t'_\rank) * \psi *
  \Asterisk_{j=1}^m q_j(t^j_1, \ldots, t^j_{\#q_j})\), where $\psi$ is
  a separating conjunction of equational atoms. By the definition of
  characteristic formul{\ae}, we obtain \(\charform{u} \equiv \exists
  y_1 \ldots \exists y_n ~.~ t_1 \mapsto (t'_1,\ldots,t'_\rank) * \psi
  * \Asterisk_{j=1}^m \charform{\proj{u}{j}}\) and, since
  $(\astore,\aheap) \models \charform{u}$, there exist locations
  $\ell_1, \ldots, \ell_n \in \locs$ and heaps $\aheap_0, \ldots,
  \aheap_m$ such that $\aheap = \biguplus_{j=0}^m \aheap_j$,
  $(\astore',\aheap_0) \models t_1 \mapsto (t'_{1},\ldots,t'_{\rank})
  * \psi$ and $(\astore',\aheap_j) \models \charform{\proj{u}{j}}$,
  for all $j \in \interv{1}{m}$, where $\astore' \isdef
  \astore[y_1\leftarrow\ell_1,\ldots,y_n\leftarrow\ell_n]$.  By the
  induction hypothesis, there exist embeddings $\plab_j$ of
  $\proj{u}{j}$ into $\aheap_j$, for each $j \in
  \interv{1}{m}$. Moreover, the sets $\nodes(\proj{u}{j})$ are
  pairwise disjoint, for all $j \in \interv{1}{m}$ and $\dom(\aheap_0)
  \isdef \set{\astore'(t_1)}$, because $(\astore',\aheap_0) \models
  t_1 \mapsto (t_{i_1}, \ldots, t_{i_\rank}) * \psi$. We define the
  mapping: \[\plab \isdef \set{(\lambda,\astore'(t_1))} \cup
  \bigcup_{j=1}^m \{(jw,\plab_j(w)) \mid w \in \nodes(\proj{u}{j})\}\]
  Clearly $\plab$ is a bijection between $\nodes(u) = \set{\lambda}
  \cup \bigcup_{j=1}^m \set{jw \mid w \in \nodes(\proj{u}{j})}$ and
  $\dom(h) = \set{\astore'(t_1)} \cup \bigcup_{j=1}^m
  \dom(\aheap_j)$. To show that $\plab$ is an embedding of $u$ into
  $\aheap$, let $wi \in \nodes(u)$ be a node, for some $i \in
  \nat$. We distinguish the following cases: \begin{compactitem}
  \item If $w = jv$, for some $j \in \interv{1}{m}$, then $vi \in
    \nodes(\proj{u}{j})$ hence $\plab_j(vi) \in \aheap_j(\plab_j(v))$,
    by the induction hypothesis, hence $\plab(wi) \in
    \aheap(\plab(w))$.
  \item Otherwise, $w = \lambda$ and $\plab(w) = \astore'(t_1)$, by the
    definition of $\plab$. Since $wi \in \nodes(u)$, it must be the
    case that $i \in \interv{1}{m}$, since, by Definition
    \ref{def:unfolding-tree}, the only children of the root of $u$ are
    $1,\ldots,m$. We have $\aheap(\plab(w)) = (\astore'(t'_1), \ldots,
    \astore'(t'_\rank))$, because $(\astore',\aheap_0) \models t_1
    \mapsto (t'_1,\ldots,t'_\rank) * \psi$ and $\aheap =
    \biguplus_{j=0}^m \aheap_j$. By the induction hypothesis, we have
    $\plab_i(\lambda) = \astore'(t^i_1)$ and, since $\asys$ is
    connected, we have $t^i_1 \in \{t_1, \ldots, t_{\rank}\}$, hence
    $\plab_i(\lambda) \in \aheap(\plab(w))$. By construction, we have
    $\plab(i) = \plab_i(\lambda) \in \aheap(\plab(w))$, thus
    concluding the proof.
  \end{compactitem}
\end{compactitem}
}}

The embedding whose existence is stated by Lemma \ref{lemma:labeling}
provides a way of \emph{decorating} the allocated locations in a heap
by the predicate symbols that caused their allocation. Given a
structure $(\astore,\aheap)$ such that $(\astore,\aheap) \models_\asys
p(t_1,\ldots,t_{\#p})$ a \emph{\labeling of $\aheap$
  w.r.t.\ $p(t_1,\ldots,t_{\#p})$} is a function $\psymfunc :
\dom(\aheap) \rightarrow \preds$ defined as $\psymfunc(\ell) \isdef q$
if and only if $u(\plab^{-1}(\ell))=(q(s_1,\ldots,s_{\#q}),\psi)$, for
some unfolding tree $u \in \utrees{p(t_1,\ldots,t_{\#p})}{\asys}$ such
that $(\astore,\aheap) \models \charform{u}$ and some embedding
$\plab$ of $u$ into $\aheap$.
Note that the unfolding tree $u$ and function $\plab$ are not unique,
hence {\labeling}s are not unique in general.

The third condition  ensuring decidability requires that all the
existentially quantified variables introduced during an unfolding can
only be associated with locations that are allocated in the heap of
any model of the formula \comment[np]{added:} (this condition is equivalent to the one given in \cite{IosifRogalewiczSimacek13}).


\begin{definition}\label{def:establishment}
  A set of rules $\asys$ is \emph{established} if and only if, for
  each rule $p(x_1, \ldots, x_{\#p}) \Leftarrow_\asys \exists z_1
  \ldots \exists z_m ~.~ \psi$ and for each $\asys$-model
  $(\astore,\aheap)$ of $\psi$, we have $\astore(z_1), \ldots,
  \astore(z_m) \in \dom(\aheap)$.
\end{definition}
Checking establishment is co-\np-hard
\cite{JansenKatelaanMathejaNollZuleger17}. \comment[np]{is there any upper bound?}
In the following, we consider only sets of rules that are progressing,
connected and established (PCE). The interest for PCE sets of rules
is motivated by the following decidability result, proved in
\cite{IosifRogalewiczSimacek13}:

\comment[np]{referee 2 made the following remark: ``This theorem could be strengthened by considering the number of quantifier alternations in [8]''. I do not understand this. Any idea?} 
\begin{theorem}\label{thm:decidability}
  Given a PCE set of rules $\asys$ and two formul{\ae} $\phi$ and
  $\psi$ the problem $\phi \models_\asys \psi$ belongs to \elementary.
  \comment[np]{one could replace this by 2EXPTIME, if this is proven in the LPAR paper?}
\end{theorem}
The rest of this paper is concerned with proving that the entailment
problem $\phi \models_\asys \psi$, for PCE sets of rules $\asys$, is
\twoexptime-hard. Previously, an \exptime-hard lower bound for this
problem was established in \cite{DBLP:conf/atva/IosifRV14}.

\section{Alternating Turing Machines} 

The proof of \twoexptime-hardness relies on a reduction from the
membership problem for alternating Turing machines. We recall some
basic definitions below.

\begin{definition}
An {\em Alternating Turing Machine} (\atm) is a tuple $M =
(Q,\Gamma,\delta,q_0,g)$ where: \begin{compactitem}
\item $Q$ is a finite set of control {\em states},
\item $\Gamma=\set{\gamma_1, \ldots, \gamma_N, \blank}$ is a finite
  alphabet, $\blank$ is the \emph{blank} symbol,
\item $\delta \subseteq Q \times \Gamma \times Q \times (\Gamma
  \setminus \set{\blank}) \times \set{\leftmv, \rightmv}$ is the
  \emph{transition relation}, $(q,a,q',b,\mu) \in \delta$ meaning
  that, in state $q$, upon reading symbol $a$, the machine moves to
  state $q'$, writes $b\neq\blank$ to the tape\footnote{A machine
    never writes blank symbols, that are used only for the initially
    empty tape cells.} and moves the head by one to the left [resp.\ right]
  if $\mu =\ \leftarrow$ [resp.\ $\mu =\ \rightarrow$],
\item $q_0 \in Q$ is the {\em initial state}, and
\item $g : Q \rightarrow \set{\vee, \wedge}$ partitions the set of
  states into \emph{existential} ($g(q) = \vee$) and \emph{universal}
  ($g(q)=\wedge$) states.
\end{compactitem}
\end{definition}

A \emph{configuration} of an ATM $M = (Q,\Gamma,\delta,q_0,g)$ is a
tuple $(q,w,i)$ where $q \in Q$ is the current state, $w : \nat
\rightarrow \Gamma$ represents the contents of the \emph{tape} and is
such that $\card{\set{j \in \nat \mid w(j) \neq \blank}} < \infty$,
and $i \in \interv{0}{\max{\set{j \in \nat \mid w(j) \neq \blank}}+1}$
is the current position of the head on the tape. We denote by
$\epsilon$ the empty word over $\Gamma$.  For any tape $w$ and integer
$i$, we denote by $\rep{w}{i}{a}$ the tape $w'$ such that $w'(i) = a$
and $w'(j) = w(j)$ for all $j \neq i$. In the following, we write \comment[np]{deleted: $w_i\isdef w(i)$,} 
$\mv{i}{\leftmv} \isdef i-1$ if $i > 0$
($\mv{0}{\leftmv}$ is undefined) and $\mv{i}{\rightmv} \isdef i+1$,
respectively. Note that, since $0$ denotes the leftmost position on
the tape, no transition moves the head left of $0$. 

The \emph{step relation} of $M$ is the following relation between
configurations: $(q,w,i) \step{(q,a,q',b,\mu)} (q',w',j)$ if and only
if there exists a transition $(q,a,q',b,\mu) \in \delta$ such that
$w(i) = a$, $w'=w[i\leftarrow b]$ and $j=i^\mu$ is defined, i.e., either
$i > 0$ or $\mu \not = \leftmv$. We omit specifying the transition
$(q,a,q',b,\mu)$ when it is not important. An \emph{execution} is a
sequence $(q_0,w_0,0) \step{(q_0,a_0,q_1,b_0,\mu_0)} (q_1,w_1,i_1)
\step{(q_1,a_1,q_2,b_1,\mu_1)} \ldots$ Note that an execution is
entirely determined by the initial configuration $(q_0,w_0,0)$ and the
sequence $(q_0,a_0,q_1,b_0,\mu_0), (q_1,a_1,q_2,b_1,\mu_1), \ldots$ of
transition rules applied to it.

Given a function $f : \nat \rightarrow \nat$, an execution is \comment[np]{important modif}
\emph{$f$-space bounded} if and only 
$\len{w_i} \leq f(\len{w_0})$, for all $i >
0$. The \atm\ $M$ is \emph{exponential-space bounded} if there exists  a constant $c$ such that every execution is 
$f$-space bounded, where $f(x)=c \cdot 2^{g(x)}$ for some constant $c$ and some univariate polynomial function 
$g$.

\begin{definition}\label{def:derivation}
  A \emph{derivation} of an \atm\ $M = (Q,\Gamma,\delta,q_0,g)$, 
  starting from a configuration $(q_0,w_0,0)$, is a \comment[np]{added:} finite tree $t$, whose
  nodes are either: \begin{compactenum}
  \item \emph{branching nodes} labeled with configurations $(q,w,i)
    \in Q \times \Gamma^*\times\nat$, or
  \item \emph{action nodes} labeled with tuples $(a,b,\mu) \in \Gamma
    \times \Gamma \setminus \set{\blank} \times
    \set{\leftarrow,\rightarrow}$, where $a$ is the symbol read, $b$
    is the symbol written and $\mu$ is the move of the head at that
    step,
  \end{compactenum}
  such that the root of $t$ is a branching node
  $t(\lambda)=(q_0,w_0,0)$ and, moreover: \begin{compactenum}[a.]
  \item\label{it1:derivation} each branching node labeled by $(q,w,i)$
    such that $g(q)=\vee$ has exactly one child, which is an
    action node labeled by $(a,b,\mu)$, where $(q,a,q',b,\mu) \in
    \delta$; the child of which is a branching node labeled by
    $(q',w',j)$, such that $(q,w,i) \step{(q,a,q',b,\mu)} (q',w',j)$;
  \item\label{it2:derivation} each branching node labeled by $(q,w,i)$
    such that $g(q)=\wedge$ has exactly one child for each tuple
    $(q,a,q',b,\mu) \in \delta$ such that $a = w(i)$; this child
    is an action node labeled by $(a,b,\mu)$, the child of which
    is a branching node labeled by $(q',w',j)$, such that $(q,w,i)
    \step{(q,a,q',b,\mu)} (q',w',j)$.
  \end{compactenum}  
  We say that $M$ \emph{accepts} $w$ if and only if $M$ admits a
  derivation starting from $(q_0,w,0)$.
\end{definition}
Note that the leaves of the tree are necessarily branching nodes
labeled by a triple $(q,w,i)$ such that $g(p)=\wedge$ and there is no
transition $(q,a,q',b,\mu)$ with $a = w(i)$.

\begin{figure}[htb]
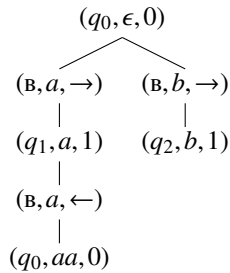

\Tree [.{$(q_0,\epsilon,0)$}
		[.{$(\blank,a,\rightmv)$}
			[.{$(q_1,a,1)$}
				[.{$(\blank,a,\leftmv)$}
					{$(q_0,aa,0)$}
				]
			]
		]
		[.{$(\blank,b,\rightmv)$}
					{$(q_2,b,1)$}
		]
 ]
	\caption{Derivation of \atm\ in Example \ref{ex:atm}}
        \label{fig:deriv}
        \vspace*{-\baselineskip}
 \end{figure}

\begin{example}
  \label{ex:atm}
Consider an \atm\ $M = (Q,\Gamma,\delta,q_0,g)$, where: $Q = \{
q_0,q_1,q_2 \}$, $\Gamma = \{ a,b,c,\blank)$, $\delta = \{
(q_0,\blank,a,q_1,\rightmv), (q_0,\blank,b,q_2,\rightmv),
(q_0,b,b,q_2,\rightmv), (q_0,b,b,q_1,\rightmv),
(q_0,c,c,q_2,\rightmv), (q_1,\blank,a,q_0,\leftmv) \}$, $g(q_0) =
q(q_2) = \wedge$ and  $g(q_1) = \vee$.  A derivation for $M$,
starting from an empty tape $\epsilon$, is depicted in Figure
\ref{fig:deriv} (the ATM contains additional transitions not used
here, they will be useful in upcoming examples). The run is on a tape
of length $2$, hence the position is encoded by a single digit.
\hfill$\blacksquare$
\end{example}

\begin{definition}\label{def:membership}
The \emph{membership problem} $(M,w)$ asks the following: given an
\atm\ $M=(Q,\Gamma,\delta,q_0,g)$ and a word
$w\in(\Gamma\setminus\set{\blank})^*$ does $M$ accept $w$ ?
\end{definition}

The complexity class \aexpspace\ is the class of membership problems
where $M$ is exponential-space bounded. It is known
that \aexpspace= \coaexpspace= \twoexptime\ \cite{ChandraKozenStockmeyer81},
where \coaexpspace\ is the complement class
of \aexpspace\footnote{Every \atm can be complemented in linear time,
  by interchanging the existential with the universal states, thus all
  alternating classes are closed under complement.}.

In the following, we shall consider only the membership problem
$(M,\epsilon)$. This is without loss of generality; indeed, let
$(M,w)$ be any instance of the membership problem, and let $c$ and $g$
be the constant and polynomial function witnessing the fact that $M$
is exponential-space bounded. Let $M_w$ be an \atm\ that produces $w$
starting from input $\epsilon$. Clearly, $M_w$ uses at most $\len{w}$
working space, thus the machine $M_w;M$, which runs $M_w$ on the empty
word and then continues with $M$, runs in space $c\cdot
2^{g(\len{w})}$ and accepts $\epsilon$ if and only if $M$ accepts
$w$. If $\nn \geq \log_2(c) + g(w)$, then $M_w;M$ runs in space
$2^{\nn}$ and moreover, $(M,w)$ and $(M_w,\epsilon)$ have the same
answer.  Therefore, we assume from now on that $M =
(Q,\Gamma,\delta,q_0,g)$ is an \atm\ started in the configuration
$(q_0,\epsilon,0)$ and that $M$ runs in space at most $2^\nn$ on the
empty input word, where $\nn$ is bounded by a polynomial in the length
of $w$.

\section{The Reduction}

This section describes the reduction of the membership problem
(Definition \ref{def:membership}) for exponential-space bounded {\atm}s
to the entailment problem (see Definition \ref{def:entailment}) for PCE sets of rules (Definitions
\ref{def:progress-connectivity} and \ref{def:establishment}). The main
idea of the reduction is the following. Since the membership problem
is existential (asking for the existence of a derivation) and the
entailment problem is universal (every model of the left-hand side is
a model of the right-hand side), a direct reduction is not
possible. Instead, we reduce from the complement of the membership
problem $(M,w)$ (there is no derivation of $M$ on $w$) to an
entailment problem instance $p_M(x) \models_{\asys_M} c_M(x)$, where
$p_M(x)$, $c_M(x)$ are predicate atoms and $\asys_M$ is a PCE set of
rules derived from the description of $M$. Intuitively, $p_M(x)$
defines all heaps with a \labeling that simulates the control structure of
$M$ (i.e.\ the branching and action nodes alternate and the control
states given by the \labeling are consistent with the transitions of
$M$), with no regard to the tape contents or the position of the
head. Then $c_M(x)$ defines only those heaps that encode derivations
violating the correctness of some tape contents or that of some
position of the head. Consequently, $p_M(x) \models_{\asys_M} c_M(x)$
holds if and only if $M$ has no derivation on $w$. Since $M$ is space
bounded by $2^\nn$, where $\nn$ is bounded by a polynomial in the
length of the input word $w$, we reduce from an arbitrary
co-\aexpspace\ problem to the entailment problem for PCE sets of
rules. Because co-\aexpspace= \aexpspace= \twoexptime, we obtain the
lower bound on the entailment problem for PCE sets of rules.

\subsection{Syntactic Shorthands}
\label{sec:shorthands}

Before giving the definitions of $p_M$, $c_M$ and $\asys_M$, we
introduce several syntactic shorthands that simplify the presentation.
 To simplify notations, we shall
assume in the remainder of the paper that all heaps and unfolding
trees are defined on the extended syntax.  For instance, although the
final encoding uses only binary heaps, i.e.\ for $\rank=2$, we shall
actually write formul{\ae} in which points-to atoms refer to arbitrary
tuples, with the convention that these tuples are always encoded as
binary heaps. More precisely, we shall write $\aheap(\ell) =
(\ell_1,\ell_2,\ell_3)$ to state that $\ell$ refers to a pair
$(\ell_1,\ell_1')$ where $\ell_1'$ itself refers to $(\ell_2,\ell_3)$,
and this additional location $\ell_1'$ \comment[np]{modif} will never be explicitly referred to. 
  Similarly, unfolding trees will also be
defined by taking into account this syntactic extension, i.e.,
points-to atoms with arbitrary tuples will be allowed to occur in the
labels of the unfolding trees, bearing in mind that such atoms will
actually yield additional unfolding steps, which will not be
explicitly considered in the tree.


\vspace*{\baselineskip}
\paragraph{Encoding Tuples}
Let $\vec{t} = (t_1,\dots,t_n)$ be a tuple of terms, with $n > 2$.
Let $\psi = \psi_1 * \dots * \psi_n$ be a (possibly empty) separated
conjunction of predicate atoms, where the first argument of every
predicate atom in $\psi_i$ is $t_i$.  By writing:
\[p(\vec{x}) \Leftarrow \exists y_1 \ldots \exists y_r ~.~ 
x_1 \mapsto (t_1, \ldots, t_n) * \psi\]
we denote the rules:
\[
\begin{array}{rcl}
p(\vec{x}) & \Leftarrow & \exists z_1 \exists y_1 \ldots \exists y_r ~.~ 
x_1 \mapsto (t_1,z_1) * \psi_1 * \widetilde{p}_1(z_1,\vec{x},y_1,\dots,y_r) \\
\widetilde{p}_i(z_i,\vec{x},y_1,\dots,y_r) & \Leftarrow &
\exists z_{i+1}~.~ z_i \mapsto (t_{i+1},z_{i+1}) * \psi_{i+1} * \widetilde{p}_{i+1}(z_{i+1},\vec{x},y_1,\dots,y_r) 
\text{, for $i \in \interv{1}{n-2}$} \\
\widetilde{p}_{n-2}(z_{n-1},\vec{x},y_1,\dots,y_r) & \Leftarrow &
z_{n-1} \mapsto (t_{n-1},t_{n}) * \psi_n \\
\end{array}
\]
where $\widetilde{p}_1,\dots,\widetilde{p}_{n-1}$ are fresh pairwise distinct predicate symbols.

The intuition is that the tuple $(t_1,\dots,t_n)$ is represented by a
binary tree of the form $(t_1,(\dots,(t_{n-1},t_n)\dots))$ of depth
$n-1$.  This allows one to encode records of various, non constant
lengths $n$ by using only a constant number of record fields (i.e.,
$\rank = 2$).  Note that the obtained rules are progressing, and, by
definition of $\psi_1,\dots,\psi_n$, they are connected. \comment[np]{added:} 
Moreover they are established (if the initial rule is established), since the variables
$y_1,\dots,y_r$ are allocated in $\psi$ and every variable $z_i$ is allocated by $\widetilde{p}_i$.    In the
following the term $(s^n,\vec{t})$ will be a shorthand for
$(\underbrace{s,\dots,s}_{\text{$n$ times}},\vec{t})$ and
$[\vec{t}]^n$ will stand for $(\nil^n,\vec{t})$.  The interest of such
special tuples will be explained later (essentially we
need to introduce ``dummy'' cells $(\nil,\dots,\nil)$ to ensure that
the rules are progressing).

%
%

\vspace*{\baselineskip}
\paragraph{Global Variables}
We assume the existence of the following \emph{global} variables that
occur free in each formula: $\vec{0}, \vec{1}, \gamma_1, \ldots,
\gamma_N$. The variables $\vec{0}$ and $\vec{1}$ denote binary digits,
and the variable $\gamma_i$ ($1 \leq i \leq N$) denote non-blank
symbols from the alphabet $\Gamma$\footnote{Since any membership
  problem is equivalent to a membership problem on a binary alphabet,
  via a binary encoding of $\Gamma$, having just $\vec{0}$ and
  $\vec{1}$ suffices. We consider distinct alphabet symbols
  $\gamma_1,\ldots,\gamma_N$ only to avoid clutter.}. These variables
will always be assigned pairwise distinct allocated locations, as
required by the following rules: \newcommand{\nnn}{a}
\begin{eqnarray}
  \allcst(x) & \Leftarrow & x \mapsto (\vec{0}, \vec{1}, \gamma_1, \ldots, \gamma_N) * 
\nnn(\vec{0}) * \nnn(\vec{1}) * {\!\Asterisk}_{i=1}^N \nnn(\gamma_i) \label{eq:const} \\
\nnn(x) & \Leftarrow & x \mapsto (\nil,\nil) \label{eq:nn}
\end{eqnarray}
Considering global variables is without loss of generality in the
following, because these variables can be added to the parameter list
of each \goal in the system (at the expense of cluttering the
presentation).

\vspace*{\baselineskip}
\paragraph{Binary Choices}
We introduce a special symbol $\blind$ which, when occurring in the body of a
rule, ranges over the global variables $\vec{0}$ and $\vec{1}$. Thus
any rule of the form:
\[
p(x_1,\ldots,x_{\#p}) \Leftarrow \exists z_1 \ldots \exists z_n ~.~
x_1 \mapsto (\blind, y) * \psi
\]
stands for the following two rules:
\[\begin{array}{rcl}
p(x_1,\ldots,x_{\#p}) & \Leftarrow & 
\exists z_1 \ldots \exists z_n ~.~ x_1 \mapsto (\vec{0}, y) * \psi \\
p(x_1,\ldots,x_{\#p}) & \Leftarrow & 
\exists z_1 \ldots \exists z_n ~.~ x_1 \mapsto (\vec{1}, y) * \psi \\
\end{array}\]
and similarly for rules of the form $p(x_1,\ldots,x_{\#p}) \Leftarrow
\exists z_1 \ldots \exists z_n ~.~ x_1 \mapsto (y,\blind) * \psi$.
The elimination of the occurrences of $\blind$ must be done {\em
  after} the encoding of tuples by binary trees, so that the number of
rules is increased by a constant $\rank^2 = 2^2$. Note also that the
fact that each rule allocates only one cell and that $\rank = 2$ (more
generally that $\rank$ is a constant) is essential here, since
otherwise the elimination of $\blind$ would yield an exponential
blow-up.
\begin{example}
\label{ex:blindelim}	
A rule $p(x) \Leftarrow x \mapsto (\blind^4)$ is first transformed
into: {\small \[
\begin{tabular}{lll}
$p(x)\Leftarrow \exists x_1~.~ x \mapsto (\blind,x_1) * p_1(x_1)$ & 
$p_1(x_1) \Leftarrow \exists x_2~.~ x_1 \mapsto (\blind,x_2) * p_2(x_2)$ &
$p_2(x_2) \Leftarrow x_2 \mapsto (\blind,\blind)$ \\
\end{tabular}
\]}
Afterwards, the symbol $\blind$ is eliminated, yielding: {\small \[
\begin{tabular}{llllll}
$p(x)$ & $\Leftarrow$ $\exists x_1~.~ x \mapsto (\vec{0},x_1) * p_1(x_1)$ 
& $p(x)$ & $\Leftarrow$ $\exists x_1~.~ x \mapsto (\vec{1},x_1) * p_1(x_1)$ \\
$p_1(x_1)$ & $\Leftarrow$ $x_1 \mapsto (\vec{0},x_2) * p_2(x_2)$ 
& $p_1(x_1)$ & $\Leftarrow$ $x_1 \mapsto (\vec{1},x_2) * p_2(x_2)$ \\
$p_2(x_2)$ & $\Leftarrow$ $x_2 \mapsto (\vec{0},\vec{0})$ 
& $p_2(x_2)$ & $\Leftarrow$ $x_2 \mapsto (\vec{0},\vec{1})$ \\
$p_2(x_2)$ & $\Leftarrow$ $x_2 \mapsto (\vec{1},\vec{0})$ 
& $p_2(x_2)$ & $\Leftarrow$ $x_2 \mapsto (\vec{1},\vec{1})$ \\
\end{tabular}
\]}
We obtain $4*2 = 8$ rules. If the first transformation is omitted then
we get $2^4= 16$ rules. \hfill$\blacksquare$
\end{example}

\vspace*{\baselineskip}
\paragraph{Binary Variables}
A binary variable $b$ is understood as ranging over the domain of the
interpretation of $\vec{0}$ and $\vec{1}$, namely the locations
assigned to $\vec{0}$ and $\vec{1}$ by the formula $\allcst$
(\ref{eq:const}). Additionally, for each binary variable $b$, we
consider the associated variable $\overline{b}$, intended to denote
the complement of $b$. More precisely, the formula $\exists b ~.~
\psi$ is to be understood as $\psi[\vec{0}/b,\vec{1}/\overline{b}] \vee
\psi[\vec{1}/b,\vec{0}/\overline{b}]$. However, this direct
substitution of the (existentially quantified) binary variables by
$\vec{0}$ and $\vec{1}$ within the rules of an established system
would break the establishment condition (Definition
\ref{def:establishment}), because $\vec{0}$ and $\vec{1}$ are not
necessarily allocated within the body of the rule\footnote{In fact
  they are allocated by the side condition $\allcst$.}. This problem
can be overcome by passing $\vec{0}$ and $\vec{1}$ as parameters to a
fresh predicate. More precisely, a rule of the form (with $1 \leq i \leq m$):
\begin{equation}\label{eq:binary-quant-rule}
p(x_1,\ldots,x_{\#p}) \Leftarrow \exists b_1 \ldots \exists b_i
\exists y_1 \ldots \exists y_n ~.~ x_1 \mapsto [\vec{t}]^m * \psi
\end{equation}
 is a shorthand for the following set of rules:
\[\begin{array}{rcl}
p(x_1,\ldots,x_{\#p}) & \Leftarrow & \exists y ~.~ x_1 \mapsto (\nil,y) * p'(y,x_1,\ldots,x_{\#p},\vec{0}, \vec{1}) \\
p(x_1,\ldots,x_{\#p}) & \Leftarrow & \exists y ~.~ x_1 \mapsto (\nil,y) * p'(y,x_1,\ldots,x_{\#p},\vec{1}, \vec{0}) \\
p'(y,x_1,\ldots,x_{\#p},b_1,\overline{b}_1) & \Leftarrow & \exists b_2 \ldots \exists b_i \exists y_1 \ldots \exists y_n ~.~ 
y \mapsto [\vec{t}]^{m-1} * \psi
\end{array}\]
Clearly, the elimination of the binary existential quantifiers from
the rule (\ref{eq:binary-quant-rule}) adds $2\cdot i$ rules to the
set. Note that the hat $[\vec{t}]^m$, of height $m \geq i$ decreases
at each step of the elimination \comment[np]{modifs} which ensures that the definition is well-founded.
 It is easy to check that the resulting rules  are progressing, connected and established. 
The rule
(\ref{eq:binary-quant-rule}) is equivalent to $2^i$ rules of the form
$p(x_1,\ldots,x_{\#p}) \Leftarrow \exists y_1 \ldots \exists y_n ~.~
x_1 \mapsto [\vec{t}]^m * \psi$ where every $b_j$ is replaced by
$\vec{0}$ or $\vec{1}$ and $\overline{b_j}$ is replaced by the
complement of $b_j$. However, adding the variables $b_j$ and
$\overline{b_j}$ one by one as parameters to the predicate allows one
to represent these rules concisely, using only $2\cdot i$ additional
rules. This comes with a cost: since the progress condition requires
each rule to allocate exactly one location, the vector $\vec{t}$ must
be embedded into a tuple $[\vec{t}]^m$ of length at least $i$.

Next, \comment[np]{added:} we introduce a syntactic shorthand to denote 
disequality  constraints on vectors of binary variables. For a vector $\vec{b} = (b_1, \ldots, b_n)$ of binary variables,
we denote by $\overline{\vec{b}}$ the vector $(\overline{b}_1, \ldots,
\overline{b}_n)$. The following rule:
\begin{equation}\label{eq:disequality-rule}
p(x_1, \ldots, x_{\#p}) \Leftarrow \exists c_1 \ldots \exists
c_n\exists y_1 \ldots \exists y_m ~.~ x_1 \mapsto \vec{t} * \psi \mid
(c_1, \ldots, c_n) \not\approx \overline{(b_1, \ldots, b_n)}
\end{equation}
where each $c_i$ ($1 \leq i \leq n$) occurs at most \comment[np]{instead of only (re-check)} once in $\vec{t}$ and
does not occur in $\psi$ and $b_1, \ldots, b_n \in \{ x_1, \ldots,
x_{\#p}\}$, is a shorthand for the following set of rules:
\begin{equation}
\label{eq:disequality-rule-decode}
  p(x_1, \ldots, x_{\#p}) \Leftarrow \exists y_1 \ldots \exists y_m
  ~.~ x_1 \mapsto \left(\vec{t}[b_i/c_i]\right)[\blind/c_j]_{j \in
    \interv{1}{n} \setminus \set{i}} * \psi \text{, $i \in
    \interv{1}{n}$}
\end{equation}
Intuitively, the rule (\ref{eq:disequality-rule}) introduces new
binary variables $c_1, \ldots, c_n$, such that not all of them are
equal to the complement of $b_1, \ldots, b_n$, respectively. In other
words, one $c_i$ must be equal to $b_i$, for some $i \in
\interv{1}{n}$, and the other $c_j$ for $j \not = i$ are arbitrary
(hence they can be replaced by $\blind$ since they occur only once in
$\vec{t}$). Note that expanding rule (\ref{eq:disequality-rule}) as
described above (\ref{eq:disequality-rule-decode}) results in at most $n$ rules of the form
(\ref{eq:binary-quant-rule}), hence the full elimination of binary
variables from the system is possible in polynomial
time.  This is mainly because in our reduction,
described next, both $i$ \comment[np]{modifs}  (in (\ref{eq:binary-quant-rule})) and $n$ (in (\ref{eq:disequality-rule})) are bounded by $\nn$, which in
turn, is polynomially bounded by the length of the input to the
membership problem.

\subsection{Pseudo-derivations as Heaps}

In this section, we show how to encode the general structure of a
derivation as a heap and define a set of rules that generates exactly
the structures corresponding to these derivations.  Importantly, since
$M$ starts on the empty word $\epsilon$, the tape contents in a
branching node can be derived from the sequence of actions along the
path from the root to that node. For this reason, we shall not
explicitly represent tape contents within the configurations and
simply label branching nodes with pairs $(q,i) \in Q \times
\interv{0}{2^\nn-1}$. We first define \emph{pseudo}-derivations, in
which the conditions on derivations are relaxed by removing all the
constraints related to the content of the tape and the position of the
head (such conditions will be considered in Section \ref{sect:deriv}).
In other words, in a pseudo-derivation, the \atm\ is treated as a mere
alternating automaton, enriched with arbitrary (possibly inconsistent)
read/write/move actions on the tape. More formally:

\begin{wrapfigure}{R}{0.3\textwidth}
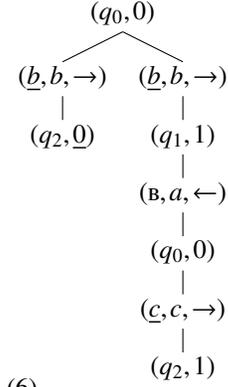

\Tree [.{$(q_0,0)$}
		[.{$(\underline{b},b,\rightmv)$}
		  {$(q_2,\underline{0})$}
                ]
		[.{$(\underline{b},b,\rightmv)$}
		  [.{$(q_1,1)$}
		    [.{$(\blank,a,\leftmv)$}
		      [.{$(q_0,0)$}
			[.{$(\underline{c},c,\rightmv)$}
			  {$(q_2,1)$}
			]
		      ]
		    ]
		  ]
		]
]
\caption{Pseudo-derivation of \atm\ in Example \ref{ex:atm}}
\label{fig:wrongpderiv}
\vspace*{-\baselineskip}
\end{wrapfigure}

\begin{definition}\label{def:pderivation}
  A \emph{pseudo-derivation} of $M = (Q,\Gamma,\delta,q_0,g)$ is a tree $t$, whose nodes are
  either: \begin{compactenum}
  \item \emph{branching nodes} labeled with pairs $(q,i)
    \in Q \times \nat$, or
  \item \emph{action nodes} labeled with tuples $(a,b,\mu) \in \Gamma
    \times \Gamma \setminus \set{\blank} \times
    \set{\leftarrow,\rightarrow}$, where $a$ is the symbol read, $b$
    is the symbol written and $\mu$ is the move of the head at that
    step,
  \end{compactenum}
  such that the root of $t$ is a branching node,
  $t(\lambda)=(q_0,0)$ and, moreover: \begin{compactenum}[a.]
  \item\label{it1:pderivation} each branching node labeled by $(q,i)$,
    such that $g(q)=\vee$, has exactly one child, that is an
    action node labeled by $(a,b,\mu)$, where $(q,a,q',b,\mu) \in
    \delta$, whose child is a branching node labeled by $(q',j)$
    such that $(q,a,b,q',\mu) \in \delta$ and $j\in \nat$;
  \item\label{it2:pderivation} each branching node labeled by $(q,i)$
    where $g(q)=\wedge$ has exactly one child for each tuple
    $(q,a,q',b,\mu) \in \delta$; this child is an action node
    labeled by $(a,b,\mu)$, the child of which is a branching node
    labeled by $(q',j)$, where $j\in \nat$.
  \end{compactenum}  
\end{definition}
\comment[np]{added:} Definition \ref{def:pderivation} is similar to Definition \ref{def:derivation} except that all the conditions related to the content of the tape and to the position of the head have been  removed (i.e., 
one does not check that the symbol $a$ occurs at position $i$ in the tape or that $j = i^\mu$).
Any derivation starting from an empty tape $\epsilon$ can be
associated with a pseudo-derivation, simply by replacing the label
$(q,w,i)$ of the branching nodes by $(q,i)$.  Conversely, for some
pseudo-derivations, we may obtain an isomorphic derivation by
inductively replacing the labels of the branching nodes from the root
to the leaves as follows.  Initially, the label $(q_0,0)$ of the root
of the tree is replaced by $(q_0,\epsilon,0)$.  Afterwards, if a
branching node is relabeled by $(q,w,i)$ and is followed by an action
node $\omega$ labeled by $(a,b,\mu)$, then the label $(q',i')$ of the
branching node following $\omega$ is replaced by $(q',w[i\leftarrow
  b],i')$.  If the obtained tree is a derivation, then we say that the
pseudo-derivation {\em yields a derivation}.  Note that this is not
always the case, because the conditions on the read actions and on the
moves in the tape are not necessarily satisfied: a branching node
$(q,w,i)$ may be followed by an action $(a,b,\mu)$ such that $a \not =
w[i]$, and the latter node may be followed by a branching node
$(q',w',i')$ with $i' \not = i^\mu$.  Figure \ref{fig:wrongpderiv} 
gives an example of a pseudo-derivation yielding no derivation, for
the ATM of Example \ref{ex:atm}. The parts of the labels that do not
fulfill the desired properties are underlined (the symbols
$\underline{b}$ and $\underline{c}$ do not match the symbols read on
the tape, and $\underline{0}$ does not match the position of the
head). The conditions ensuring that a pseudo-derivation yields a
derivation will be given in Section \ref{sect:deriv}.

We represent the pseudo-derivations of $M$ as tree-shaped heaps
generated by a set of rules where, intuitively, each predicate $q(x)$
allocates a branching node labeled by a pair $(q,i)$ and each
predicate $\act{q}(x,a,b,\mu)$ allocates an action node labeled
$(a,b,\mu)$. \comment[np]{added:} In our representation, the state $q$ will actually be 
omitted (see, e.g., Rule (\ref{rule:branch-exists})), because it is implicitly defined by 
the unfolding
tree. 
 Further, we represent each position $i \in \interv{0}{2^\nn-1}$ on
the tape succintly, by an $\nn$-tuple of binary digits $\bin{i} \in
\set{\vec{0}, \vec{1}}^\nn$ and encode the left and right moves as
$\movenc{\leftarrow} \isdef \vec{0}$ and $\movenc{\rightarrow} \isdef
\vec{1}$. Let $\transet{q,a} \isdef \delta \cap \left(\set{q} \times
\{a \} \times Q \times \Gamma \setminus \set{\blank} \times
\set{\leftarrow,\rightarrow}\right)$ be the set of transitions of $M$
with source state $q$, reading symbol $a$ from the tape. We consider
the following rules, for each state $q \in Q$ and symbol $a \in
\Gamma$:
\begin{eqnarray}
q(x) & \Leftarrow & \exists x' ~.~ x \mapsto (\blind^\nn,x') * \act{q}'(x',a,b,\movenc{\mu}) \label{rule:branch-exists}
\\[-1mm]
&& \text{ if $g(q)=\vee$ and $(q,a,q',b,\mu) \in \transet{q,a}$} \nonumber\\[1mm]
 q(x) & \Leftarrow & \exists y_1 \ldots \exists y_n ~.~ x \mapsto (\blind^\nn,y_1, \ldots, y_m) * 
 \Asterisk_{j=1}^m ~\act{q}_j(y_j,a,b_j,\movenc{\mu}_j) \label{rule:branch-univ}
\\[-1mm]
&&\text{ if $g(q)=\wedge$ and $\transet{q,a} = \{(q,a,q_1,b_1,\mu_1), \ldots,(q,a,q_m,b_m,\mu_m)\}$} \nonumber\\[1mm]
\act{q}(x,y,z,u) & \Leftarrow & \exists x' ~.~ x \mapsto (y, z, u, x') * q(x') \label{rule:action}
\end{eqnarray}
The heaps defined by the above rules ensure only that the control
structure of a derivation of $M$ is respected, namely that the
branching and action nodes alternate correctly, and that the sequence
of control states labeling the branching nodes on any path is
consistent with the transition relation of $M$. In other words, these
trees encode pseudo-derivations of $M$. Further, we introduce a
top-level predicate $p_M(x)$ that allocates the special variables
$\vec{0}, \vec{1}, \gamma_1, \ldots, \gamma_N$ and ensures that the
initial state $q_0$ of $M$ is the first control state that occurs on
an path of a pseudo-derivation:
\begin{eqnarray}
  p_M(x) & \Leftarrow & \exists y \exists z ~.~ x \mapsto (y, z) * p_M'(y) * \allcst(z)
  \label{rule:pM} \\
  p_M'(y) & \Leftarrow & \exists z'~.~ y \mapsto [z']^\nn * q_0(z')
  \label{rule:ppM}
\end{eqnarray}
The hat $[z']^\nn$ above ensures that every heap generated by $p_M'$
begins with a tuple $[z']^\nn \isdef
(\overbrace{\nil,\dots,\nil}^\nn,z')$. The use of this tuple will be
made clear in Section \ref{sect:deriv}. For now, let $\asys_M$ be the
set consisting of the rules above.  In the
following, we stick to the convention that predicate symbol $q$
represents a branching node, whereas $\act{a}$ represents an action
node. The definition below formalizes the encoding of a
pseudo-derivation by a structure:

\newcommand{\principal}{principal}

\comment[np]{in def below, $\aheap$ replaced by $\aheap_2$ in items $3$ and $4$ (equivalent)}

\begin{definition}\label{def:pseudo-derivation-encoding}
  A structure $(\astore,\aheap)$ such that $(\astore,\aheap)
  \models_{\asys_M} p_M(x)$ \emph{encodes} a pseudo-derivation $t$ of
  $M$, written as $\encodes{(\astore,\aheap)}{}{t}$, if and only if
  there exists a \labeling $\psymfunc$ of $\aheap$ w.r.t.\ $p_M(x)$,
  two heaps $\aheap_1$ and $\aheap_2$ and a bijection $f : \nodes(t)
  \rightarrow \dom(\aheap_2)$ such that, for all $w \in \nodes(t)$,
  the following hold: \begin{compactenum}
  \item \label{it0:pseudo-derivation-encoding} $\aheap = \aheap_1
    \uplus \aheap_2$,
  \item \label{it1:pseudo-derivation-encoding} $(\astore,\aheap_1)
    \models \exists y \exists z \exists z' ~.~ x \mapsto (y, z) *
    \allcst(z) * y \mapsto [z']^\nn$,
  \item\label{it2:pseudo-derivation-encoding} If $w$ is a branching
    node with label $t(w)=(q,i)$ and children $w0,\dots,wn$,
    then $\psym{}{f(w)}=q$ and $\aheap_2(f(w)) = (\ell_1, \ldots,
    \ell_{\nn},f(w0),\dots,f(wn))$, where $\ell_j =
    \astore(\bin{i}_j)$, for all $j \in \interv{1}{\nn}$, 
  \item\label{it3:pseudo-derivation-encoding} If $w$ is an action node
    with label $t(w)=(a,b,\mu)$ and only child $w0$, then we have
    $\aheap_2(f(w)) = (\astore(a), \astore(b), \astore(\movenc{\mu}),
    f(w0))$.
  \end{compactenum}
\end{definition}

\newcommand{\atree}{u}
\newcommand{\atreehat}{\hat{u}}
\newcommand{\atreetop}{\atree'}

\begin{figure}
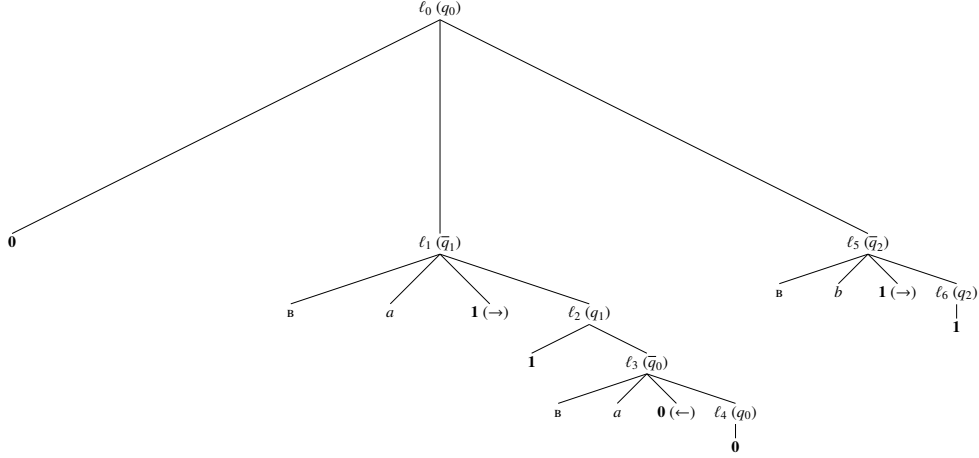

  \begin{center}
    \scalebox{0.6}{
      \Tree [.{$\ell_0\ (q_0)$}
	$\vec{0}$  
	[.{$\ell_1\ (\act{q}_1)$} 
	  $\blank$
	  $a$
	  $\vec{1}\ (\rightmv)$
	  [.{$\ell_2\ (q_1)$}
	    $\vec{1}$
	    [.{$\ell_3\ (\act{q}_0)$}
	      $\blank$
	      $a$ 
	      $\vec{0}\ (\leftmv)$	
	      [.{$\ell_4\ (q_0)$}
		$\vec{0}$
	      ]
	    ]
	  ]
	]
	[.{$\ell_5\ (\act{q}_2)$}
	  $\blank$
	  $b$
	  $\vec{1}\ (\rightmv)$
	  [.{$\ell_6\ (q_2)$}
	    $\vec{1}$
	  ]
	]
    ]}
    \caption{A heap encoding the derivation of Figure \ref{fig:deriv}}
    \label{fig:hderiv}
  \end{center}
\end{figure}

A heap encoding the derivation of Figure \ref{fig:deriv} is depicted
in Figure \ref{fig:hderiv} (for readability, the part corresponding to
the formula $\exists y \exists z \exists u ~.~ x \mapsto (y, z) *
\allcst(z) * y \mapsto [z']^\nn$ is not depicted, i.e., only the heap
$\aheap_2$ of Definition \ref{def:pseudo-derivation-encoding} is
shown).  We also give, for each location $\ell$, the corresponding
predicate $\psym{}{\ell}$. 

\begin{lemma}\label{lemma:pseudo-derivation}
  \begin{inparaenum}[(A)]
    \item\label{it1:pseudo-derivation} For each pseudo-derivation $t$
      of $M$, 
      there exists a
      structure $(\astore,\aheap) \models_{\asys_M} p_M(x)$ such that
      $\encodes{(\astore,\aheap)}{}{t}$.
    \item\label{it2:pseudo-derivation} Dually, for each structure
      $(\astore,\aheap) \models_{\asys_M} p_M(x)$, there exists a
      pseudo-derivation $t$ of $M$ such that
      $\encodes{(\astore,\aheap)}{}{t}$.
  \end{inparaenum}
\end{lemma}
\shortversion{}{
\proof{
(\ref{it1:pseudo-derivation}). Let $t$ be a pseudo-derivation of
$M$. We first build an unfolding tree $\atree$ as follows.  We let
$\nodes(\atree)\isdef\nodes(t)$ and associate to each $w\in \nodes(t)$
a variable $x_w\notin \set{\vec{0}, \vec{1}, \gamma_1,\ldots,
  \gamma_N}$ such that $w\neq w'\Rightarrow x_w\neq w_{w'}$.  We shall
define $\atree$ such that $\atree \in
\utrees{q_0(x_\lambda)}{\asys_M}$. Consider a branching node $w\in
\nodes(t)$ where $t(w) = (q,i)$; we define the label of $w$ and its
children as follows, depending on $g(q)$.
\begin{compactitem}
\item If $g(q) = \vee$, then $w0$ [resp.\ $w00$] is the only child of
  $w$ [resp.\ $w0$], with labels $t(w0)=(a, b, \mu)$ and $t(w00)=(q',
  i')$. We then define:
  \[\begin{array}{rcll}
  \atree(w) & \isdef & \left(q(x_w),~ \exists x_{w0} ~.~ x_w \mapsto (\bin{i},x_{w0}) * \act{q'}(x_{w0},a,b,\movenc{\mu})\right) & ~\text{(rule \ref{rule:branch-exists})} \\
  \atree(w0) & \isdef & \left(\act{q'}(x_{w0}, a, b, \movenc{\mu}),~ \exists x_{w00} ~.~ x_{w0} \mapsto (a, b, \movenc{\mu}, x_{w00}) * q'(x_{w00}) \right) & ~\text{(rule \ref{rule:action})}
  \end{array}\]
\item Otherwise, $g(q) = \wedge$, the nodes $wj$ are the children of
  $w$ and the unique child of $wj$ is $wj0$, with labels $t(wj)=(a,
  b_j, \mu_j)$ and $t(wj0)=(q'_j, i'_j)$, for all $j \in
  \interv{0}{n}$. We then define:
  \[\begin{array}{rcll}
  \atree(w) & \isdef & \left(q(x_w),~ \exists x_{w0} \ldots \exists x_{wn} ~.~ x \mapsto (\bin{i},x_{w0}, \ldots, x_{wn}) * 
  \Asterisk_{j=0}^n \act{q}_j(x_{wj},a,b_j,\movenc{\mu}_j)\right) & ~\text{(rule \ref{rule:branch-univ})} \\
  \atree(wj) & \isdef & \left(\act{q}_j(x_{wj}),~ \exists x_{wj0} ~.~ x_{wj} \mapsto (a, b_j, \movenc{\mu}_j, x_{wj0}) * q_j(x_{wj0})\right) & ~\text{(rule \ref{rule:action})}
  \end{array}\]
\end{compactitem}
Next, $u$ is extended into an unfolding tree $\atreehat \in
\utrees{p_M(x)}{{\asys_M}}$, defined as follows (where $x,y,z$ are
pairwise distinct variables not occurring in $\set{\vec{0}, \vec{1},
  \gamma_1, \ldots, \gamma_N}$ and distinct from the variables
associated with the nodes in $t$): {\footnotesize
	\[
	\Tree 
	[.{$(p_M(x), \exists y \exists z ~.~ x \mapsto (y, z) * p_M'(y) * \allcst(z))$} 
	[ .{$(p_M'(y), \exists x_\lambda~.~ y \mapsto [x_\lambda]^\nn * q_0(x_\lambda))$}
	{$\atree$}
	]
	[.{$(\allcst(z), \nnn(\vec{0}) * \nnn(\vec{1}) *
		{\!\Asterisk}_{i=1}^N \nnn(\gamma_i))$}
	{$(\nnn(\vec{0}), \vec{0} \mapsto (\nil,\nil))$}
	{$(\nnn(\vec{1}), \vec{1} \mapsto (\nil,\nil))$}
	{$(\nnn(\vec{\gamma_i}), \gamma_i \mapsto (\nil,\nil))$}
	]
	]
	\]          }                          
It is clear that $\atreehat \in \utrees{p_M(x)}{{\asys_M}}$ and that
$\charform{\atreehat} = \exists y \exists z \exists x_\lambda ~.~ x \mapsto (y, z) * \allcst(z) * y \mapsto [x_\lambda]^\nn * \charform{u}$ is satisfiable, since there are no equality or
disequality atoms and all nodes allocate distinct variables. 
Thus there exists a heap $\aheap$ such that 
$(\astore,\aheap) \models \charform{\atreehat}$, and we deduce that  there exist $\aheap_1, \aheap_2$ such that
conditions (\ref{it0:pseudo-derivation-encoding}) and
(\ref{it1:pseudo-derivation-encoding})  from Definition
\ref{def:pseudo-derivation-encoding} are satisfied,  with $\aheap_1$ denoting the part of the heap such that $(\astore,\aheap_1) \models \exists y \exists z \exists x_\lambda ~.~ x \mapsto (y, z) * \allcst(z) * y \mapsto [x_\lambda]^\nn$ and $\aheap_2$ denoting the part of the heap such that $(\astore,\aheap_2)  \models \exists x_\lambda~.~ \charform{u}$.
To check
that $\encodes{(\astore,\aheap)}{}{t}$, we need to exhibit a
bijection $f : \nodes(t) \rightarrow \dom(\aheap_2)$ that
meets conditions 
(\ref{it2:pseudo-derivation-encoding}) and
(\ref{it3:pseudo-derivation-encoding}) from Definition
\ref{def:pseudo-derivation-encoding}.
 Because ${\asys_M}$ is a progressing and
connected set of rules and $(\astore,\aheap_2) \models
\charform{\atree}$, by Lemma \ref{lemma:labeling}, there exists an
embedding $f$ of $\atree$ into $\aheap_2$ and Points
(\ref{it2:pseudo-derivation-encoding}) and
(\ref{it3:pseudo-derivation-encoding}) follow straightforwardly from
the definition of $\atree$ above.

(\ref{it2:pseudo-derivation}). If $(\astore,\aheap_2)
\models_{\asys_M} p_M(x)$ then, by the definition of
$\models_{\asys_M}$, there exists an unfolding tree $u \in
\utrees{p_M(x)}{\asys_M}$ such that $(\astore,\aheap) \models
\charform{u}$.  By definition of the rules in $\asys_M$, we have
$\charform{u} = \exists y \exists z \exists x_\lambda ~.~ x \mapsto
(y, z) * \allcst(z) * y \mapsto [x_\lambda]^\nn *
\charform{\proj{u}{00}}$, where $\proj{u}{0}$ is labeled by $(p_M'(y),
\exists x_\lambda~.~ y \mapsto [x_\lambda]^\nn * q_0(x_\lambda))$ and
$\proj{u}{00}$ is labeled by a pair of the form
$(q_0(x_\lambda),\phi)$.  Thus there exist $\aheap_1,\aheap_2$ such
that $\aheap = \aheap_1 \uplus \aheap_2$, $(\astore,\aheap_1) \models
\exists y \exists z \exists x_\lambda ~.~ x \mapsto (y, z) *
\allcst(z) * y \mapsto [x_\lambda]^\nn$ and $(\astore',\aheap_2)
\models_{\asys_M} q_0(x_\lambda)$, for some extension $\astore'$ of
$\astore$.  It is straightforward to check that $u'=\proj{u}{00}$
%
%
is an unfolding tree.  Since $\asys_M$ is progressing and connected,
by Lemma \ref{lemma:labeling}, there exists an embedding $\Lambda$ of $u'$ into $\aheap_2$. 
We build a pseudo-derivation $t$ of $M$ such that $\nodes(t) =
\nodes(u')$, by induction on the structure of $u'$.  Note that, since
each $\gamma \in \Gamma \cup \set{\vec{0}, \vec{1}}$ is allocated
separately in $\charform{u}$, the restriction of $\astore$ to the set
$ \Gamma \cup \set{\vec{0}, \vec{1}}$ is a bijection.  For each $w \in
\nodes(u')$: \begin{compactitem}
\item If $u'(w)$ is of the form $(q(x),\phi)$ then $\phi$ is the body
  of rule (\ref{rule:branch-exists}) or (\ref{rule:branch-univ}). In
  case (\ref{rule:branch-exists}), we have $\aheap(\Lambda(w)) =
  (\ell_1, \ldots, \ell_{\nn+1})$, with $\ell_1,\ldots,\ell_\nn \in
  \set{\astore(\vec{0}),\astore(\vec{1})}$. Let $i$ be the integer
  such that $\bin{i} =
  (\astore^{-1}(\ell_1),\ldots,\astore^{-1}(\ell_\nn))$; we set $t(w)
  \isdef (q,i)$. The construction for case (\ref{rule:branch-univ}) is
  handled analogously.
\item Otherwise, by definition of $\asys_M$, necessarily $u'(w)$ is of
  the form $(\act{q}(x, a, b, \movenc{\mu}),\phi)$ and $\phi$ is the
  body of rule (\ref{rule:action}). In this case, we have
  $\aheap(\Lambda(w)) = (\ell_1, \ell_2, \ell_3, \ell_4)$, with
  $\ell_1 \in \astore(\Gamma)$, $\ell_2 \in \astore(\Gamma \setminus
  \set{\blank})$ and $\ell_3 \in \set{\astore(\vec{0}),
    \astore(\vec{1})}$. In this case, we define $t(w) \isdef
  (\astore^{-1}(\ell_1), \astore^{-1}(\ell_2), \mu)$, where
  $\mu=\leftarrow$ if $\ell_3=\astore(\vec{0})$ and $\mu=\rightarrow$
  if $\ell_3=\astore(\vec{1})$.
\end{compactitem}
It is easy to check that $t$ is a pseudo-derivation of
$M$. To verify that $\encodes{(\astore,\aheap)}{}{t}$, we take
$f : \nodes(t) \rightarrow \dom(\aheap_2)$ as the function $\Lambda$. 
Clearly, $f$ is a bijection and the conditions 
(\ref{it2:pseudo-derivation-encoding}) and (\ref{it3:pseudo-derivation-encoding}) of Definition
\ref{def:pseudo-derivation-encoding} are straightforward checks. 

  }}
  

\subsection{Encoding Complement Membership as Entailment Problems}
\label{sect:deriv}

In this section, we show how to encode the conditions that ensure that
a pseudo-derivation is a derivation, namely that the considered
pseudo-derivation also fulfills all the conditions related to the tape
contents and the position of the head.  More precisely, we recall that
a pseudo-derivation of $M$ yields a derivation of $M$ if the contents
of the tape and the head's position are consistent with the sequence
of actions leading to that particular configuration. This is the case
if the following conditions hold: \begin{compactenum}[I.]
\item\label{cond:dep} If a branching node labeled $(q,i)$ is followed
  by an action node labeled $(a,b,\rightmv)$ [resp.\ $(a,b,\leftmv)$], itself
  followed by a branching node labeled $(q',i')$ then necessarily $i'
  = i+1$ [resp.\ $i = i'+1$], i.e.\ the position of the head changes
  according to the action executed between the adjacent configurations (for instance, in Figure \ref{fig:wrongpderiv}, the
  position $\underline{0}$ does not fulfill this condition). 
\item\label{cond:tape} For every $i \in \interv{0}{2^\nn-1}$, if along
  a path from a branching node labeled $(q,i)$ followed by an action
  node labeled $(a,b,\mu)$, to another branching node labeled
  $(q',i)$ followed by an action node labeled $(a',b',\mu')$, there
  is no branching node labeled $(q'',i)$, then necessarily $a'=b$.
  Indeed, the symbol read on position $i$ must be the one previously
  written, since it was not changed in the meantime (e.g., in Figure \ref{fig:wrongpderiv}, the
  symbol $\underline{c}$ does not fulfill this condition).
\item\label{cond:emptyinit} For every $i \in \interv{0}{2^\nn-1}$, if
  along a path from the root to a branching node labeled $(q,i)$,
  followed by an action node labeled $(a,b,\amove)$, there is no
  branching node labeled $(q',i)$, then necessarily $a = \blank$,
  i.e.\ the tape is initially empty (e.g., this condition is violated by the symbol $\underline{b}$ in Figure \ref{fig:wrongpderiv}).
\end{compactenum}

In the following, we shall not check that the above conditions hold
for some derivation of $M$, but rather the opposite: that for each
pseudo-derivation of $M$, at least one of the above conditions is
broken. In other words, we reduce from the complement of the
membership problem $(M,\epsilon)$ to an entailment problem, defined
next. This does not change the final \twoexptime-hardness result,
because, as previously mentioned, \twoexptime= \aexpspace= \coaexpspace.

To this end, we consider a predicate $c_M$ and a set of rules
$\asys_M$ containing rules for $p_M(x)$ and $c_M(x)$ such that the
entailment $p_M(x) \models_{\asys_M} c_M(x)$ holds if and only if
every pseudo-derivation of $M$ violates at least one of the conditions
(\ref{cond:dep}), (\ref{cond:tape}) or (\ref{cond:emptyinit}); in
other words, if and only if $M$, started on input $\epsilon$, admits
no derivation.

Let $\nt \isdef \max_{q \in Q, a \in \Gamma} \card{\transet{q,a}}$ be
the maximum branching degree (i.e.\ the maximum number of children of
a node) of a derivation of $M$. We define an auxiliary predicate
$r(x)$ that generates all tree-shaped heaps in which branching nodes
correctly alternate with action nodes, with no regard for the labels
of those nodes:
\[\begin{array}{rcl}
r(x) & \Leftarrow & \exists y_1 \ldots \exists y_n ~.~ x \mapsto (\blind^\nn, y_1, \ldots, y_n) * \Asterisk_{j=1}^n \act{r}(y_j) 
\text{, for each $n \in \interv{0}{\nt}$} \\
\act{r}(x) & \Leftarrow & \exists y ~.~ x \mapsto (a, b, \blind, y) * r(y)
\text{, for each $a \in \Gamma$ and $b \in \Gamma \setminus \set{\blank}$}
\end{array}\]

First, we define the heap encodings of those pseudo-derivation trees
that violate condition (\ref{cond:dep}). To this end, we guess a
vector $\vec{b}$ in $\{0,1\}^\nn$, encoding a position on the tape $i
\in \interv{0}{2^\nn-1}$, a shift $\amove \in \set{\leftmv,\rightmv}$,
encoded by $\movenc{\amove} \in \set{\vec{0},\vec{1}}$ and get the
binary complement of the (encoding of the) position reached from
$\vec{b}$ by applying $\amove$. Here we distinguish two cases,
depending on the choice of $\amove$: \begin{compactenum}[(a)]
\item\label{it1:choice} If $\amove$ is $\rightmv$ then we guess $\bin{i} =
  \vec{b} \isdef (b_1,\dots, b_n, \vec{0}, \vec{1}^{\nn-1-n})$ for some
  $n\in \interv{0}{\nn-1}$ and let $\vec{c} \isdef (\overline{b}_1,
  \ldots, \overline{b}_n, \vec{0}, \vec{1}^{\nn-1-n})$ be the
  complement of $\bin{i+1}=(b_1, \dots, b_n, \vec{1},
  \vec{0}^{\nn-1-n})$.
\item\label{it2:choice} Otherwise, $\bin{i} = \vec{b} \isdef (b_1,
  \dots, b_n, \vec{1}, \vec{0}^{\nn-1-n})$ and let $\vec{c} \isdef
  (\overline{b}_1, \ldots, \overline{b}_n, \vec{1},
  \vec{0}^{\nn-1-n})$ be the complement of $\bin{i-1} = (b_1, \dots,
  b_n, \vec{0}, \vec{1}^{\nn-1-n})$.
\end{compactenum}
For every $n \in \interv{0}{\nn-1}$,  $m \in \interv{0}{\nt}$
and $i \in \interv{1}{m}$, we consider the following rules: 
{\small
\begin{eqnarray}
c_1(x) & \Leftarrow & \exists b_1\! \ldots\! \exists b_n \exists y ~.~ x \mapsto \left([y]^\nn\right) * 
d_1(y,\vec{1},\underbrace{b_1\ldots b_n,\vec{0},\vec{1}^{\nn-n-1}}_{\vec{b}}, 
\underbrace{\overline{b}_1\ldots\overline{b}_n,\vec{0},\vec{1}^{\nn-n-1}}_{\vec{c}})
\label{rule:choiceI1} \\
c_1(x) & \Leftarrow & \exists b_1\! \ldots\! \exists b_n \exists y ~.~ x \mapsto \left([y]^\nn\right) * 
d_1(y,\vec{0},\underbrace{b_1\ldots b_n,\vec{1},\vec{0}^{\nn-n-1}}_{\vec{b}}, 
\underbrace{\overline{b}_1\ldots\overline{b}_n,\vec{1},\vec{0}^{\nn-n-1}}_{\vec{c}})
\label{rule:choiceI2}
\end{eqnarray}

\begin{eqnarray}
d_1(x,u,\vec{b},\vec{c}) & \Leftarrow & \exists y_1 \ldots \exists y_m ~.~ x \mapsto (\blind^\nn,y_1, \ldots, y_m) * 
\!\!\!\!\!\!\!
\Asterisk_{j \in \interv{1}{m} \setminus \set{i}}
\!\!\!\!\!\!\!\!\!\!\!
\act{r}(y_j) * \act{d}_1(y_i,u,\vec{b},\vec{c})
\label{rule:branch-pathI} \\
d_1(x,u,\vec{b},\vec{c}) & \Leftarrow & \exists y_1 \ldots \exists y_m ~.~ x \mapsto (\vec{b},y_1, \ldots, y_m) *
\!\!\!\!\!\!\!
\Asterisk_{j \in \interv{1}{m} \setminus \set{i}}
\!\!\!\!\!\!\!\!\!\!\!
\act{r}(y_j) * \act{e}_1(y_i,u,\vec{b},\vec{c})
\label{rule:branch-nodeI} \\
\act{d}_1(x,u,\vec{b},\vec{c}) & \Leftarrow & \exists y ~.~ x \mapsto (a,b,\blind,y) * d_1(y,u,\vec{b},\vec{c})
\text{, for each $a \in \Gamma$, $b \in \Gamma \setminus \set{\blank}$}
\label{rule:action-pathI} \\
\act{e}_1(x,u,\vec{b},\vec{c}) & \Leftarrow & \exists y ~.~ x \mapsto (a,b,u,y) * f_1(y,\vec{b},\vec{c})
\text{, for each $a \in \Gamma$, $b \in \Gamma \setminus \set{\blank}$}
\label{rule:action-nodeI} \\
f_1(x,\vec{b},\vec{c}) & \Leftarrow &
\exists y_1 \ldots \exists y_m \exists \vec{e} ~.~ x \mapsto (\vec{e},y_1, \ldots, y_m) * \Asterisk_{j=1}^m \act{r}(y_j)
\mid \vec{e} \not\approx \overline{\vec{c}}
\label{rule:branch-violatesI}
\end{eqnarray}}

\begin{figure}[htb]
  \begin{center}
    \begin{minipage}[t][][b]{0.49\textwidth}
      \centerline{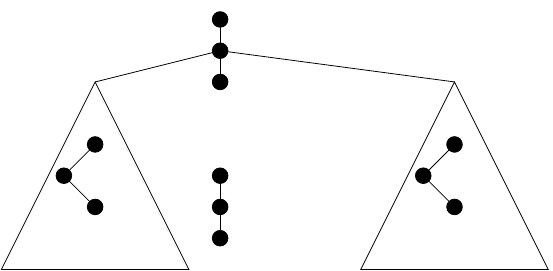}
      \centerline{(\ref{cond:dep})}
    \end{minipage}
    \begin{minipage}{0.49\textwidth}
      \centerline{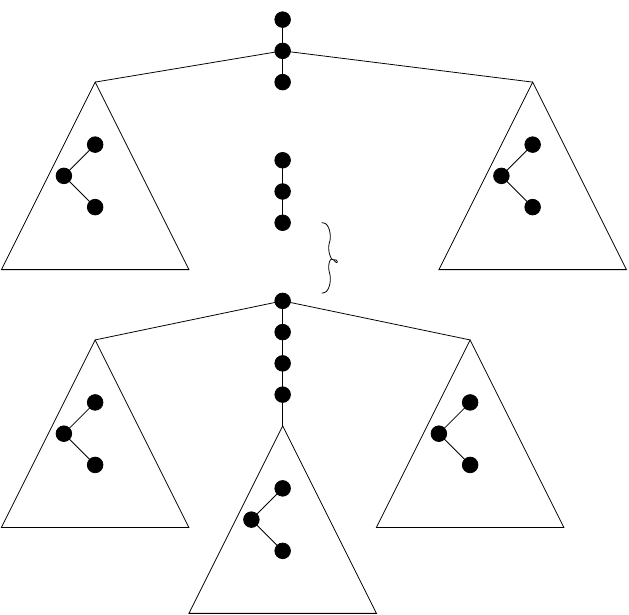}
      \centerline{(\ref{cond:tape})}
    \end{minipage}
  \end{center}
\caption{Pseudo-derivations violating conditions (\ref{cond:dep}) and
  (\ref{cond:tape})} \label{fig:cond}
\end{figure}

For a graphical depiction of the idea behind the encoding of
violations of condition (\ref{cond:dep}), we refer to Figure
\ref{fig:cond} (\ref{cond:dep}). Intuitively, rules
(\ref{rule:choiceI1}) and (\ref{rule:choiceI2}) choose the move
$\mu\in\set{\leftmv,\rightmv}$ (encoded by $\vec{0}$ or $\vec{1}$) and
the binary vectors $\vec{b},\vec{c} \in \set{\vec{0},\vec{1}}^\nn$,
according to the cases (\ref{it1:choice}) and (\ref{it2:choice})
above, respectively. Note that we use the hat $[y]^\nn$ to eliminate
the binary variables $b_1, \ldots, b_n$, as $n < \nn$, according to
the elimination procedure described in \S\ref{sec:shorthands}. Then a
path to the branching node, labeled $(q',i')$, that violates condition
(\ref{cond:dep}) is non-deterministically  chosen, by alternating the branching and action
nodes allocated by rules (\ref{rule:branch-pathI}) and
(\ref{rule:action-pathI}), respectively. The offending branching node
is allocated by rule (\ref{rule:branch-violatesI}) and its
predecessors are the branching and the action nodes, labeled with
$(q,i)$ and $(a,b,\mu)$, such that $i' \neq i^\mu$. These latter nodes
are allocated by rules (\ref{rule:branch-nodeI}) and
(\ref{rule:action-nodeI}), respectively.

The pseudo-derivations of $M$ that violate condition (\ref{cond:tape})
are encoded by the tree-structured heaps defined by the rules
below. To this end, we guess a binary vector $\vec{b} \in
\set{\vec{0}, \vec{1}}^\nn$ denoting the position of a write action
that has an inconsistent read descendant and let $\vec{c}$ be its
binary complement. Then, for every $m \in \interv{0}{\nt}$ and $i \in
\interv{1}{m}$, we consider the rules below \comment[np]{added:} (explanations will be provided later):
\begin{eqnarray}
c_2(x) & \Leftarrow & \exists b_1 \ldots \exists b_{\nn} \exists y ~.~ x \mapsto \left([y]^\nn\right) * 
d_2(y,\underbrace{b_1,\ldots,b_{\nn}}_{\vec{b}},\underbrace{\overline{b}_1,\ldots,\overline{b}_{\nn}}_{\vec{c}}) 
\label{rule:choiceII} \\[-2mm]
d_2(x,\vec{b},\vec{c}) & \Leftarrow & \exists y_1 \ldots \exists y_m ~.~ x \mapsto (\blind^\nn, y_1,\ldots,y_m) *
\!\!\!\!\!\!\!
\Asterisk_{j \in \interv{1}{m} \setminus \set{i}}
\!\!\!\!\!\!\!\!\!\!\!
\act{r}(y_j) * \act{d}_2(y_i,\vec{b},\vec{c})
\label{rule:branch-path-beforeII} \\
\act{d}_2(x,\vec{b},\vec{c}) & \Leftarrow & \exists y ~.~ x \mapsto (a, b, \blind, y) * d_2(y, \vec{b}, \vec{c})
\text{, for each $a \in \Gamma$, $b \in \Gamma \setminus \set{\blank}$}
\label{rule:action-path-beforeII} \\
\act{d}_2(x,\vec{b},\vec{c}) & \Leftarrow & \exists y ~.~ x \mapsto (a, b, \blind, y) * e_2(y, \gamma, \vec{b}, \vec{c})
\text{, for each $a \in \Gamma$, $\gamma \in \Gamma \setminus  \{ b \}$}
\label{rule:action-node-startII} \\
e_2(x,\gamma,\vec{b},\vec{c}) & \Leftarrow & \exists y_1 \ldots \exists y_m ~.~ 
x \mapsto (\vec{b},y_1, \ldots, y_m) *
\!\!\!\!\!\!\!
\Asterisk_{j \in \interv{1}{m} \setminus \set{i}}
\!\!\!\!\!\!\!\!\!\!\!
\act{r}(y_j) * \act{f}_2(y_i,\gamma,\vec{b},\vec{c}) 
\label{rule:branch-path-afterII} \\
\act{f}_2(x,\gamma,\vec{b},\vec{c}) & \Leftarrow & \exists y ~.~ x \mapsto (a, b, \blind, y) * f_2(y, \gamma, \vec{b}, \vec{c})
\text{, for each $a \in \Gamma$, $b \in \Gamma \setminus \set{\blank}$}
\label{rule:action-path-afterII} \\
\act{f}_2(x,\gamma,\vec{b},\vec{c}) & \Leftarrow & \exists y ~.~ x \mapsto (a, b, \blind, y) * g_2(y, \gamma, \vec{b}, \vec{c})
\text{, for each $a \in \Gamma$, $b \in \Gamma \setminus \set{\blank}$}
\label{rule:action-node-endII} \\
f_2(x,\gamma,\vec{b},\vec{c}) & \Leftarrow & \exists y_1 \ldots \exists y_m \exists \vec{e} ~.~ 
x \mapsto (\vec{e},y_1, \ldots, y_m) *
\!\!\!\!\!\!\!\!\!
\Asterisk_{j \in \interv{1}{m} \setminus \set{i}}
\!\!\!\!\!\!\!\!\!\!\!
\act{r}(y_j) * \act{f}_2(y_i,\gamma,\vec{b},\vec{c}) \mid \vec{e} \not\approx \overline{\vec{c}}
\label{rule:branch-node-pathII} \\
g_2(x,\gamma,\vec{b},\vec{c}) & \Leftarrow & \exists y_1 \ldots \exists y_m ~.~ 
x \mapsto (\vec{b},y_1, \ldots, y_m) *
\!\!\!\!\!\!\!\!\!
\Asterisk_{j \in \interv{1}{m} \setminus \set{i}}
\!\!\!\!\!\!\!\!\!\!\!
\act{r}(y_j) * \act{g}_2(y_i,\gamma) 
\label{rule:branch-node-endII} \\
\act{g}_2(x,\gamma) & \Leftarrow & \exists y ~.~ x \mapsto (\gamma,b,\blind,y) * r(y)
\text{, for each $b \in \Gamma \setminus \set{\blank}$}
\label{rule:action-violatesII}
\end{eqnarray}

For a depiction of the idea behind the encoding of violations of
condition (\ref{cond:tape}), we refer to Figure \ref{fig:cond}
(\ref{cond:tape}). Rule (\ref{rule:choiceII}) uses the hat $[y]^\nn$
to choose the tuple of binary variables $\vec{b}=(b_1,\ldots,b_\nn)$
and their complements
$\vec{c}=(\overline{b}_1,\ldots,\overline{b}_\nn)$. First, the path to
a branching node labeled by the binary position $\vec{b}$ is
non-deterministically chosen by an alternation of branching and action
nodes allocated by the the rules (\ref{rule:branch-path-beforeII}) and
(\ref{rule:action-path-beforeII}), respectively, until the node and
its predecessor are allocated by rules
(\ref{rule:branch-path-afterII}) and (\ref{rule:action-node-startII}),
respectively. We also guess a symbol $\gamma$, distinct from the
symbol written on the tape at position $\vec{b}$, and store it in the
second parameter of $e_2(x,\gamma,\vec{b},\vec{c})$. Next, a path to a
second branching node labeled by the binary position $\vec{b}$ is
non-deterministically chosen by an alternation of branching and action
nodes allocated by the the rules (\ref{rule:branch-node-pathII}) and
(\ref{rule:action-path-afterII}) respectively, while checking that no
branching node with the same position $\vec{b}$ occurs on this second
path (due to the side condition $\vec{e} \not\approx \overline{\vec{c}}$ of Rule (\ref{rule:branch-node-pathII}) and the fact that $\vec{b} = \overline{\vec{c}}$) . At the end, we reach the
offending branching node (\ref{rule:branch-node-endII}), whose
predecessor is allocated by rule (\ref{rule:action-node-endII}). At
this point, we check that the symbol read by the last action node is
$\gamma$ (i.e. is different than the symbol previously written at
position $\vec{b}$, by rule (\ref{rule:action-node-startII})). This
check is done by rules (\ref{rule:branch-node-endII}) and
(\ref{rule:action-violatesII}), ensuring that condition
(\ref{cond:tape}) is violated.

Next, we define the tree-structured heap encoding of the derivation
trees that violate condition (\ref{cond:emptyinit}). To this end, we
guess a binary vector $\vec{b} \in \set{\vec{0}, \vec{1}}^\nn$
denoting the position where a symbol different from $\blank$ has been
read, with no previous write action at that position and let $\vec{c}$
be its complement. We consider the rules below, for every $m
\in \interv{0}{\nt}$ and $i \in \interv{1}{m}$:  
\begin{eqnarray}
c_3(x) & \Leftarrow & \exists b_1 \ldots \exists b_{\nn} \exists y ~.~ x \mapsto ([y]^\nn) * 
d_3(y,\underbrace{b_1,\ldots,b_{\nn}}_{\vec{b}},\underbrace{\overline{b}_1,\ldots,\overline{b}_{\nn}}_{\vec{c}})
\label{rule:choiceIII} \\[-2mm]
d_3(x,\vec{b},\vec{c}) & \Leftarrow & \exists y_1 \ldots \exists y_m \exists \vec{e} ~.~ x \mapsto (\vec{e},y_1, \ldots ,y_m) *
\!\!\!\!\!
\Asterisk_{j \in \interv{1}{m} \setminus \set{i}} 
\!\!\!\!\!\!\!\!\!\!\!
\act{r}(y_j) * \act{d}_3(y_i,\vec{b},\vec{c}) \mid \vec{e} \not\approx \overline{\vec{c}}
\label{rule:branch-pathIII} \\
\act{d}_3(x,\vec{b},\vec{c}) & \Leftarrow & \exists y ~.~ x \mapsto (a,b,\blind,y) * d_3(y,\vec{b},\vec{c})
\text{, for all $a \in \Gamma$, $b \in \Gamma \setminus \set{\blank}$}
\label{rule:action-pathIII} \\
\act{d}_3(x,\vec{b},\vec{c}) & \Leftarrow & x \mapsto (a,b,\blind,y) * e_3(y,\vec{b},\vec{c})
\text{, for all $a \in \Gamma$, $b \in \Gamma \setminus \set{\blank}$}
\label{rule:action-nodeIII} \\
e_3(x,\vec{b},\vec{c}) & \Leftarrow & \exists y_1 \ldots \exists y_m ~.~ x \mapsto (\vec{b},y_1, \ldots ,y_m) *
\!\!\!\!\!
\Asterisk_{j \in \interv{1}{m} \setminus \set{i}} 
\!\!\!\!\!\!\!\!\!\!\!
\act{r}(y_j) * \act{f}_3(y_i)
\label{rule:branch-nodeIII} \\
\act{f}_3(x) & \Leftarrow & \exists y ~.~ x \mapsto (a,b,\blind,y) * r(y)
\text{, for all $a, b \in \Gamma \setminus \set{\blank}$}
\label{rule:action-violatesIII}
\end{eqnarray}
After the initial guess of the binary position $\vec{b}$, by rule
(\ref{rule:choiceIII}), a path to a branching node labeled by
$\vec{b}$ is non-deterministically guessed, by an alternation of
branching and action nodes corresponding to the rules
(\ref{rule:branch-pathIII}) and (\ref{rule:action-pathIII}),
respectively, while checking that no branching node labeled with
position $\vec{b}$ occurs on this path. Once this node is reached, by
rule (\ref{rule:action-nodeIII}), we check that its action node
child reads a symbol different than $\blank$, by rules
(\ref{rule:branch-nodeIII}) and (\ref{rule:action-violatesIII}), which
is in violation of condition (\ref{cond:emptyinit}).

Finally, the predicate $c_M(x)$ that chooses the condition
(\ref{cond:dep}), (\ref{cond:tape}) or (\ref{cond:emptyinit}) to be
violated, is defined by the following rules: 
\begin{equation}\label{rule:cM}
  c_M(x) \Leftarrow \exists y \exists z ~.~ x \mapsto (y,z) * c_i(y) * \allcst(z)
  \text{, for all $i \in \set{1,2,3}$}
\end{equation}
Let $\asys_M$ denote the set of rules introduced so far. The following
lemma states the property of the models of $c_M(x)$:

\begin{lemma}\label{lemma:violation}
  Given a pseudo-derivation $t$ of $M$ and a structure
  $(\astore,\aheap)$, such that $\actencodes{(\astore,\aheap)}{}{t}$,
  we have $(\astore,\aheap) \models_{\asys_M} c_M(x)$ if and only if $t$
  is not a derivation of $M$.
\end{lemma}
\shortversion{}{\proof{

  Since $\actencodes{(\astore,\aheap)}{}{t}$, there exist heaps
  $\aheap_1,\aheap_2$ and a bijection $f : \nodes(t) \rightarrow
  \dom(\aheap_2)$ satisfying the conditions of Definition
  \ref{def:pseudo-derivation-encoding}.
  
  ``$\Rightarrow$''. If $(\astore,\aheap) \models_{\asys_M} c_M(x)$
  then (by the definition of $\models_{\asys_M}$) there exists an
  unfolding tree $u \in \utrees{c_M(x)}{\asys_M}$ such that
  $(\astore,\aheap) \models \charform{u}$, and by definition of the
  rules in ${\asys_M}$, we have $\charform{u} = \exists y \exists z
  ~.~ x \mapsto (y, z) * \allcst(z) * \charform{\proj{u}{0}}$, where
  $u(0) = (c_i(y), \phi_i)$ for some $i\in \set{1, 2, 3}$ and some
  formula $\phi_i$. Furthermore, since ${\asys_M}$ is a progressing
  and connected set of rules, by Lemma \ref{lemma:labeling}, there
  exists an embedding $\plab$ of $u$ into $\aheap$. We
  assume that $i = 1$ and that $\phi_i$ is the body of a rule
  \ref{rule:choiceI1}; the proofs in the other cases are similar. In
  this case we have $\charform{\proj{u}{0}} = \exists y_1~.~ y
  \mapsto [y_1]^\nn * \charform{\proj{u}{00}}$, where $u(00)$ is
  labeled by $(d_1(y_1,\vec{1},\bin{j},\bin{j'})),\psi)$,
  for some $j,j'$ with $\bin{j'} =
  \overline{\bin{j+1}}$. Now, $(\astore,\aheap_1) \models \exists y
  \exists z \exists y_1 ~.~ x \mapsto (y, z) * \allcst(z) * y \mapsto
          [y_1]^\nn$ because $\actencodes{(\astore,\aheap)}{}{t}$,
          hence necessarily, $(\astore,\aheap_2) \models
          \charform{\proj{u}{00}}$.
  
  By inspection of the rules
  (\ref{rule:choiceI1})-(\ref{rule:branch-violatesI}), we conclude
  that the subtree $\proj{u}{00}$ admits a node (possibly identical to
  $00$) labeled by $(d_1(y', \vec{1}, \bin{j}, \bin{j'}), \psi')$,
  with a child node labeled by
  $(\overline{e_1}(y_\nu,\vec{1},\bin{j},\bin{j'}),\psi_\nu)$ (see
  rule \ref{rule:branch-nodeI}), the latter admitting a single child
  node labeled by $(f_1(y'',\vec{1},\bin{j},\bin{j'}),\psi'')$ (see
  rule \ref{rule:action-nodeI}). Therefore, $\charform{\proj{u}{00}}$
  contains the following points-to atoms, where $m,n\in \nat$ and
  $\nu\in \interv{1}{m}$:
  \[\begin{array}{rcll}
  y' & \mapsto & (\bin{j}, y_1, \ldots, y_m) & \text{(rule \ref{rule:branch-nodeI})}\\
  y_v & \mapsto & (a,b, \vec{1}, y'') & \text{(rule \ref{rule:action-nodeI})}\\
  y'' & \mapsto & (\vec{e}, y_1', \ldots, y_n') & \text{(rule \ref{rule:branch-violatesI})}
  \end{array}\]
  Moreover, $\vec{e}$ is of the form $\bin{k}$ with $\bin{k} \neq
  \overline{\bin{j'}}$ (see rule \ref{rule:branch-violatesI}), hence
  $k \neq i+1$.  Since $(\astore,\aheap_2) \models
  \charform{\proj{u}{00}}$, there exists an extension $\astore'$ of
  $\astore$ and locations
  $\ell',\ell_1,\dots,\ell_m,\ell'',\ell_1',\dots,\ell_n'$ such that
  $\astore'(y') = \ell'$, $\astore'(y_j) = \ell_j$ for $j \in
  \interv{1}{m}$, $\astore(y'') = \ell''$ and $\astore'(y'_j) =
  \ell'_j$ for $j\in \interv{1}{n}$; furthermore, we have
  $\aheap_2(\ell') = (\astore'(\bin{j}), \ell_1, \ldots, \ell_m)$,
  $\aheap_2(\ell_\nu) = (\astore'(a), \astore'(b), \astore'(\vec{1}),
  \ell'')$ and $\aheap_2(l'') = (\astore'(\vec{e}), \ell_1', \ldots,
  \ell_n')$. The locations $\ell',\ell_1,\dots,\ell_n',\ell''$ must
  all occur in $\dom(\aheap_2)$, which entails that $t$ contains a
  branching node $f^{-1}(\ell')$, followed by an action node
  $f^{-1}(\ell_\nu)$, itself followed by an action node
  $f^{-1}(\ell'')$, and by Definition
  \ref{def:pseudo-derivation-encoding}, we have $t(f^{-1}(\ell')) =
  (q,i)$, $t(f^{-1}(\ell_\nu)) = (a,b,\rightmv)$ and
  $t(f^{-1}(\ell'')) = (q',k)$, with $k \not = i+1$.  This contradicts
  condition (\ref{cond:dep}), thus $t$ is not a derivation of $M$.

  \noindent''$\Leftarrow$'' If $t$ is a pseudo-derivation
  but not a derivation of $M$, then $t$ violates one of the conditions
  (\ref{cond:dep}), (\ref{cond:tape}) or (\ref{cond:emptyinit}). 
  Since $(\astore,\aheap) \models p_M(x)$, 
  there exists an unfolding tree $u \in \utrees{p_M(x)}{{\asys_M}}$ such that $(\astore,\aheap) \models
  \charform{u}$. 
We then build an unfolding tree $u' \in
  \utrees{c_M(x)}{{\asys_M}}$, isomorphic to $u$, with $\charform{u} = \charform{u'}$.
    We detail the construction only for the case
where condition (\ref{cond:tape}) is violated (this is the most complex case).
In this case,  
there exist two branching nodes $w_1$ and $w_2$, in $\nodes(t)$ labeled by 
$(q_1,i)$ and $(q_2,i)$ respectively, such that: \begin{inparaenum}[(i)] 
\item{$w_2$ is below $w_1$,}
\item{
for every branching node of label $(q,j)$ 
along the path from $w_1$ to $w_2$(excluded) we have $j \not = i$,}
\item{
the child $w_1'$ of $w_1$ along the path from $w_1$ to $w_2$ 
is labeled by $(a_1,b_1,\mu_1)$,}
\item{ 
and $w_2$ has a child $w_2'$ labeled by $(a_2,b_2,\mu_2)$, with $a_2 \not = b_1$.}
\end{inparaenum} Let $\vec{b} = \bin{i}$ and $\vec{c} = \overline{\bin{i}}$.

The top of the tree $u'$ is defined as follows, in accordance to the rules defining $c_M(x)$ and $c_2(y)$:
{\footnotesize
\[
\Tree 	
			[.{$(c_M(x),x \mapsto (y, z) * c_2(y) * \allcst(z)))$}
				[.{$(c_2(y),\exists y' ~.~ y \mapsto ([y']^\nn) * 
d_2(y',\vec{b},\vec{c}))$}
					$\proj{u'}{00}$
				]
			[.{$(\allcst(z), z \mapsto (\vec{0}, \vec{1}, \gamma_1, \ldots, \gamma_N) *\nnn(\vec{0}) * \nnn(\vec{1}) *
{\!\Asterisk}_{i=1}^N \nnn(\gamma_i))$}
				{$(\nnn(\vec{0}), \vec{0} \mapsto (\nil,\nil))$}
				{$(\nnn(\vec{1}), \vec{1} \mapsto (\nil,\nil))$}
				{$(\nnn(\vec{\gamma_i}), \gamma_i \mapsto (\nil,\nil))$}
			]
			]
	\]}
The subtree $\proj{u'}{00}$ is defined as follows. We set
$\nodes(\proj{u'}{00}) = \nodes(\proj{u}{00})$ and we specify the
label $u'(w)$ of each node $w$ in $\proj{u}{00}$.  Let $w$ be such a
node. We distinguish several cases according to the position of $w$ in
$u$.  In what follows, $x'$ denotes the variable allocated at $w$ in
$u$ and $y_k$ denotes the variable allocated at the child node
$w\cdot(k-1)$ (if it exists).  Moreover, if $w$ is a node along the
path from $w_1$ to $w_2'$ but distinct from $w_2'$, then $i$ denotes
the unique $i$ such that $wi$ is a prefix of $w_2'$.  Finally, observe
that if $w$ is a branching node, $u(w)$ is necessarily of the form
$(q(\vec{e},y_1,\dots,y_m),\psi)$, for some state $q$, and if $w$ is
an action node, then $u(w)$ is of the form
$(\act{q}(a,b,\mu,y_1),\psi)$.

\begin{itemize}
\item{If $w$ is a branching node but not a prefix  of $w_2$
we set: $u'(w) = (r(x'), \exists y_1 \ldots \exists y_n ~.~ x' \mapsto (\vec{e}, y_1, \ldots, y_n) * \Asterisk_{j=1}^n \act{r}(y_j) )$.}

\item{
For any action node distinct from $w_2'$ and that is not a prefix  of $w_2$ 
we set: $u'(w) = (\act{r}(x'),  
\exists y_1 ~.~ x' \mapsto (a, b, \mu, y_1) * r(y_1))$.}

\item{If $w$ is a branching node occurring along the path between the root and $w_1$ (excluded)
we set: $u'(w) = (d_2(x',u,\vec{b},\vec{c}),\exists y_1 \ldots \exists y_m . x' \mapsto (\vec{e},y_1, \ldots, y_m) * 
\Asterisk_{j \in \interv{1}{m} \setminus \set{i}}
\act{r}(y_j) * \act{d}_2(y_i,u,\vec{b},\vec{c}))$.}

\item{If $w$ is an action node between the root and $w_1$, distinct from the predecessor of $w_1$,
we set: $u'(w) = (\act{d}_2(x',\vec{b},\vec{c}), 
\exists y_1 ~.~ x' \mapsto (a, b, \mu, y_1) * d_2(y, \vec{b}, \vec{c}))$.}

\item{If $w$ is the predecessor of $w_1$
we set: $u'(w) = (\act{d}_2(x',\vec{b},\vec{c}), \exists y_1 ~.~ x' \mapsto (a, b, \mu, y_1) * e_2(y, a_2, \vec{b}, \vec{c}))$. This fits in with the definition of the rules of $\act{d}_2$ because by definition $b$ is the symbol $b_1$ defined above and $a_2 \not = b_1$.}

\item{If $w = w_1$, then $u'(w)$ is defined as follows: \[(e_2(x',\gamma,\vec{b},\vec{c}),\exists y_1 \ldots \exists y_m 
x' \mapsto (\vec{b},y_1, \ldots, y_m) *
\Asterisk_{j \in \interv{1}{m} \setminus \set{i}} \act{r}(y_j) *  
\act{f}_2(y_i,\gamma,\vec{b},\vec{c})))\]}

\item{If $w$ is an action node between $w_1$ and $w_2$ but distinct from the predecessor of $w_2$, then 
$u'(w) = (\act{f}_2(x',\gamma,\vec{b},\vec{c}),\exists y_1 ~.~ x' \mapsto (a, b, \blind, y_1) * f_2(y_1, \gamma, \vec{b}, \vec{c}))$.}

\item{If $w$ is a branching node between $w_1$ and $w_2$ (excluded)
then we set:
\[u'(w) = (f_2(x',\gamma,\vec{b},\vec{c}), \exists y_1 \ldots \exists y_m
x' \mapsto (\vec{e},y_1, \ldots, y_m) *
\Asterisk_{j \in \interv{1}{m} \setminus \set{i}}
\act{r}(y_j) * \act{f}_2(y_i,\gamma,\vec{b},\vec{c})) \]
Note that by the above property, necessarily $\vec{e} \not = \vec{b}$ thus the 
side condition of the rule is fulfilled.
}

\item{If $w$ is the predecessor of $w_2$, then 
$u'(w) = (\act{f}_2(x',\gamma,\vec{b},\vec{c}),\exists y_1 ~.~ x' \mapsto (a, b, \mu, y_1) * g_2(y_1, \gamma, \vec{b}, \vec{c}))$.}

\item{If $w  = w_2$ then $u'(w) = 
(g_2(x',\gamma,\vec{b},\vec{c}), \exists y_1 \ldots \exists y_m
x' \mapsto (\vec{b},y_1, \ldots, y_m) *
\Asterisk_{j \in \interv{1}{m} \setminus \set{i}}
\act{r}(y_j) * \act{g}_2(y_i,\gamma))$}
 
\item{If $w  = w_2'$ then $u'(w) = 
(\act{g}_2(x',\gamma),\exists y_1 ~.~ x' \mapsto (\gamma,b,\mu,y_1) * r(y))$}

\end{itemize}
It is easy to check, by inspection of all the cases above and of the
rules in ${\asys_M}$, that $u'$ is a derivation tree, isomorphic to
$u$. Further, by construction every node in $u'$ allocates the same
heap cell than the corresponding node in $u$. Consequently,
$\charform{u'} =\charform{u}$, and $(\astore,\aheap) \models c_M(x)$.
}}


\begin{lemma}\label{lemma:entailment}
  The entailment $p_M(x) \models_{\asys_M} c_M(x)$ holds if and only
  if the membership problem $(M,\epsilon)$ has a negative answer.
\end{lemma}
\proof{ ``$\Rightarrow$'' Suppose that $M$ accepts $\epsilon$. By
  Definition \ref{def:derivation} there exists a derivation $t$
  starting from $\epsilon$.  Since $t$ is a derivation, it is also a
  pseudo-derivation of $M$ and, by Lemma
  \ref{lemma:pseudo-derivation} (\ref{it1:pseudo-derivation}), there exists a structure
  $(\astore,\aheap)$ such that $(\astore,\aheap) \models_{\asys_M}
  p_M(x)$ and $\encodes{(\astore,\aheap)}{}{t}$.  By Lemma
  \ref{lemma:violation}, we obtain $(\astore,\aheap)
  \not\models_{\asys_M} c_M(x)$, thus $p_M(x) \not\models_{\asys_M}
  c_M(x)$. ''$\Leftarrow$'' Suppose that $p_M(x) \not\models_{\asys_M}
  c_M(x)$, hence there exists a structure $(\astore,\aheap)$ such that
  $(\astore,\aheap) \models_{\asys_M} p_M(x)$ and $(\astore,\aheap)
  \not\models_{\asys_M} c_M(x)$. By Lemma
  \ref{lemma:pseudo-derivation} (\ref{it2:pseudo-derivation}), there exists a pseudo-derivation $t$
  of $M$ such that $\encodes{(\astore,\aheap)}{}{t}$. By Lemma
  \ref{lemma:violation}, $t$ is a derivation of $M$, hence
  $(M,\epsilon)$ has a positive answer. \qed}

We state the main result of this paper below:

\begin{theorem}\label{thm:complexity}
  The entailment problem $p(x) \models_\asys q(x)$, where $\asys$ is a
  progressing, connected and established set of rules and $p,q$ are
  predicate symbols in $\preds$ that occur as {\goal}s in $\asys$, is
  \twoexptime-hard.
\end{theorem}
\proof{ Given an exponential-space bounded \atm\ $M$ we define a set
  of rules $\asys_M$, based on the description of $M$, such that
  $p_M(x) \models_{\asys_M} c_M(x)$ if and only if $(M,\epsilon)$ has
  a negative answer (Lemma \ref{lemma:entailment}). Moreover, the set
  of rules is easy shown to be progressing, connected and
  established. The reduction is possible in time polynomial in the
  size of the standard encoding of $M$. Indeed, the number of rules in
  $\asys$ is $\bigO(\card{Q} \cdot \nn \cdot \nt)$ and the succint
  representation of each rule, using binary choices and binary
  variables can be generated in time $\bigO(\card{\Gamma} \cdot \nt
  \cdot \nn)$. Finally, the complete elimination of binary variables
  is possible in polynomial time. Since we reduce from the complement
  of a \aexpspace-complete problem and
  co-\aexpspace=\aexpspace=\twoexptime, we obtain the
  \twoexptime-hardness result.  \qed}

\section{Conclusion}

\comment[np]{slight modif}
The entailment problem, for symbolic heaps with
inductively defined predicates satisfying some additional conditions,
was showed to be decidable (with elementary recursive time complexity) in
\cite{IosifRogalewiczSimacek13}.
We showed that this problem has an actual \twoexptime-hard lower
bound. In the light of the recent results of
\cite{KatelaanMathejaZuleger19,ZK20,PMZ20}, this settles an open
problem concerning the tight complexity of what is currently the most
general decidable class of entailments for Separation Logic with
inductive definitions.  Note that the \twoexptime-hardness proof
relies only on entailments between atoms (more precisely they are of
the form $p(\vec{x}) \models_\asys q(\vec{x})$) and that inductive
rules defining $p$ and $q$ contain no equational atom. Further, the
constructed structures are actually quite restricted: they are
directed acyclic graphs, with ``almost'' a tree shape, where only a
polynomial number of children pointing to $(\nil,\nil)$ are shared
between nodes.  Thus, \twoexptime-hardness also holds for systems that
are restricted to generate structures of this form.  This draws a very
precise boundary for the complexity of the entailment problem in the
considered fragment of $\seplogk{\rank}$, since it is known that the
problem is \exptime-complete if the structures are trees
\cite{DBLP:conf/atva/IosifRV14} (possibly enriched with backward links
from children to parents).

Concerning future work, we are now trying to 
extend the decidability and complexity results 
to a larger class of inductive definitions, by relaxing some of the conditions in Section \ref{sect:fragment}.


\end{document}